\pdfoutput=1

\documentclass[iop]{emulateapj}

\newcommand {\Lya}    {Ly$\alpha$}   

\newcommand {\OVI}    {\ion{O}{6}}   
\newcommand {\OVII}   {\ion{O}{7}}
\newcommand {\OVIII}  {\ion{O}{8}}

\newcommand {\CIII}   {\ion{C}{3}}  
\newcommand {\CIV}    {\ion{C}{4}}

\newcommand {\NeVIII}  {\ion{Ne}{8}}

\newcommand {\NOVI}   {$N_{\rm OVI}$}

\newcommand {\dndz}  {$d{\cal N}/dz$}

\usepackage{graphics} 

\def\subsun{\mbox{$_{\odot}$}}

\shorttitle{Nature of the WHIM}
\shortauthors{Smith et al.}

\begin{document}

\title{The Nature of the Warm/Hot Intergalactic Medium I. 
  Numerical Methods, Convergence, and OVI Absorption}

\author{Britton D. Smith, 
  Eric J. Hallman\altaffilmark{1,2}, 
  J. Michael Shull}
\affil{Center for Astrophysics \& Space Astronomy,
  Department of Astrophysical \& Planetary Sciences,
  389 UCB,
  University of Colorado, Boulder, CO, 80309}

\altaffiltext{1}{National Science Foundation Astronomy and
  Astrophysics Postdoctoral Fellow}
\altaffiltext{2}{Institute for Theory and Computation,
  Harvard-Smithsonian Center for Astrophysics, Cambridge, MA 02138}

\and

\author{Brian W. O'Shea}
\affil{Department of Physics \& Astronomy,
  Michigan State University,
  East Lansing, MI 48824}

\email{britton.smith@colorado.edu, michael.shull@colorado.edu}

\begin{abstract}
We perform a series of cosmological simulations using \texttt{Enzo},
an Eulerian adaptive-mesh refinement, N-body + hydrodynamical code,
applied to study the warm/hot intergalactic medium.  The WHIM may be
an important component of the baryons missing observationally at low
redshift.  We investigate the dependence of the global star formation
rate and mass fraction in various baryonic phases on spatial
resolution and methods of incorporating stellar feedback.  Although
both resolution and feedback significantly affect the total mass in
the WHIM, all of our simulations find that the WHIM fraction peaks at $z
\sim 0.5$, declining to 35--40\% at $z = 0$.  We construct samples of
synthetic \OVI\ absorption lines from our highest-resolution
simulations, using several models of oxygen ionization balance.
Models that include both collisional ionization and photoionization provide
excellent fits to the observed number density of absorbers per unit
redshift over the full range of column densities ($10^{13}~{\rm
  cm}^{-2} \lesssim$ \NOVI\ $\lesssim 10^{15}$ cm$^{-2}$).
Models that include only collisional ionization provide better
fits for high column density absorbers (\NOVI\ $  \gtrsim
10^{14}$~cm$^{-2}$).  The distribution of \OVI\ in density and
temperature exhibits two populations: one at $T \sim 10^{5.5}$~K
(collisionally ionized, 55\% of total \OVI) and one at $T \sim
10^{4.5}$~K (photoionized, 37\%) with the remainder located  in dense
gas near galaxies.  While not a perfect tracer of hot gas, \OVI\
provides an important tool for a WHIM baryon census.
\end{abstract}

\section{Introduction}

It is well established that the fraction of baryons in the universe
that are easily observable drops from nearly 100\% at high redshift to
less than half by the current epoch.  Predictions from Big Bang Nucleosynthesis 
\citep{2000PhR...333..389O, 2007ARNPS..57..463S} 
and measurements of the cosmic microwave background by
\citet{2010arXiv1001.4538K} 
are in agreement with the cosmic baryon budget measured by observations of the \Lya\  
forest at $z > 4$.  Yet baryon surveys at $z = 0$  consistently finds
that as many as 60\% of those baryons are ``missing" 
\citep{1992MNRAS.258P..14P,1994MNRAS.267...13B, 1998ApJ...503..518F,
  2004ApJ...616..643F}.  
In the low-redshift intergalactic medium (IGM), approximately 30\% of the baryons reside
in photoionized diffuse \Lya\ absorbers (the ``Lyman-alpha forest") and $\sim10$\%
in hot gas traced by \OVI\ absorbers (Danforth \& Shull 2008).  
For reviews of the missing-baryons problem and observational attempts at a solution, 
see \citet{2007ARA&A..45..221B} and \citet{2003ASSL..281....1S}.  
Early numerical simulations \citep{1999ApJ...514....1C, 1999ApJ...511..521D, 
2001ApJ...552..473D} predicted that a majority of the missing
baryons are located in a gaseous phase of moderate to low density and
temperatures $10^{5}$--$10^{7}$~K, known as the warm/hot intergalactic medium 
(WHIM).  These simulations and many that followed showed that many baryons are 
heated into the WHIM phase via gravitational shocks from gas falling onto dark-matter 
filaments and halos, and through galactic outflows and stellar feedback.  

WHIM gas is quite tenuous and difficult to observe, with baryon
overdensities of typically 
$\delta_H \equiv \rho_{b}/\bar{\rho_{b}} \approx 0.1-100$.  Here, $\rho_{b}$ and 
$\bar{\rho_{b}}$ are the baryon density and mean baryon density, 
estimated at $\bar{\rho_{b}} \equiv \Omega_{b}\ \rho_{cr, 0}\ (1 +
z)^{3} = (4.26 \times 10^{-31}~{\rm g~cm}^{-3}) (1 + z)^{3}$, where
$\rho_{cr, 0}$ is the critical density of the universe at $z = 0$.  
The method for probing the WHIM that has proved most fruitful thus far is the 
detection of the \OVI\ doublet ($\lambda\lambda$1032, 1038) in absorption in the 
low-redshift \Lya\  forest, first accomplished by
\citet{2000ApJ...534L...1T}.  In collisional ionization equilibrium (CIE), \OVI\ reaches 
its peak abundance fraction, $f_{\rm OVI} \approx 0.2$, at $T \sim 10^{5.45}$~K
\citep{1993ApJS...88..253S}, potentially making it an ideal tracer of
WHIM gas below $10^{6}$~K.  Since the first detections, large surveys of 
low-redshift \OVI\  absorption lines have been completed
\citep{2005ApJ...624..555D, 2008ApJ...679..194D, 2006ApJ...640..716D, 
2008ApJS..177...39T, 2008ApJS..179...37T, 2008ApJ...683...22T},
providing statistical datasets for comparison with 
simulations.  \citet{2005MNRAS.359..295F} showed that \OVI\ and \OVII\
absorption statistics can be fit reasonably well by models that assume
these systems to be generated in a network of virial and infall shocks
that surround cosmological halos.  
\citet{2006ApJ...650..573C} were able to broadly reproduce the 
observed distribution in column density of \OVI\ absorbers, quantified as the
number of absorbers per unit redshift ($d{\cal N}/dz$), by including a non-equilibrium 
treatment of oxygen ionization balance and stellar feedback from
galactic superwinds.  However, they did not explicitly study whether these
absorbers are tracing the WHIM.  More recently,
\citet{2009MNRAS.395.1875O} found that a majority of 
\OVI\ absorption systems in their simulations arise from cold ($T \sim
15,000$ K) 
gas that is primarily photoionized.  They suggested that \OVI\ may not trace the 
shock-heated WHIM at all.  \citet{2005ApJ...620...21K}  arrived at a
similar conclusion.  
However, the radiative cooling method of
\citet{2009MNRAS.395.1875O} only accounted for the influence of
photoionization on the cooling of H and He, but not the metals, whose
cooling rate was calculated assuming CIE.  At moderately high
metallicities ($Z \gtrsim 0.1 Z\subsun$), this can lead to an
overestimate of the total cooling by more than an order of magnitude
for $T < 10^{5}$ K.  An excellent demonstration of this is given in
Figure 10 of \citet{2010arXiv1007.2840T}.

In this work, we present the results of a new set of numerical simulations designed to 
study the nature of the WHIM and its relation to low-redshift metal absorption systems.  
This paper is the first in a series in which we use the results of these and additional 
simulations to study the observational signatures of the WHIM.  Our
primary goals in this first paper are 
to validate our approach and provide initial predictions.  
In Section \ref{sec:methods}, we detail the numerical methods employed in this work.  
In Section \ref{sec:results}, we discuss the initial results of these simulations.  
To understand the extent to which these simulations can be trusted, we begin by 
examining how well they are converged, a topic that has received little previous
attention.  We use our highest-resolution simulations to create samples of synthetic \OVI\ 
absorption lines for comparison with observations.  We then take a closer look at the 
physical environment associated with the \OVI\ in our simulations, in
order to gauge its usefulness 
as a WHIM tracer.  We conclude with a discussion and summary in
Section \ref{sec:summary}.

\section{Methods} \label{sec:methods}

In order to understand the dependence of WHIM characteristics on the
properties of the simulations, we carry out a large number of cosmological simulations 
with varying spatial resolution, box size, and physical parameters.  In every simulation, 
the number of grid cells is equal to the number of dark-matter particles.  All 
of the simulations are initialized at $z = 99$ in a $\Lambda$CDM universe with a power spectrum of 
density fluctuations given by \citet{1999ApJ...511....5E}.  The cosmological
parameters are ($\Omega_{b}$, $\Omega_{\rm CDM}$, $\Omega_{\Lambda}$, $h$,
$\sigma_{8}$, $n$) = (0.0441, 0.2239, 0.732, 0.704, 0.82, 1), in
reasonable agreement with the WMAP-7 findings
\citep{2010arXiv1001.4538K}.  
All simulations with the same box size are initialized using identical
random seeds, and thus have the same large-scale structure, allowing
us to isolate the effects of resolution.  Simulations with different
box sizes have different random seeds.  
A summary of the simulations and their distinguishing features is given in 
Table \ref{tab:sims}.   We now describe the setup and methodology in further detail.

\begin{deluxetable*}{lcccccc}
 \tablewidth{0pt}
  \tablecaption{Parameters of Simulations}
  \tablehead{
    \colhead{Run} & \colhead{$l$} & \colhead{$N_{\rm cells}^{1/3}$}
    & \colhead{$\Delta x$} & \colhead{$m_{dm}$} 
    & \colhead{Feedback} & \colhead{Metal} \\
    \colhead{ } &  \colhead{(Mpc/$h$)} &  \colhead{ }
    & \colhead{(kpc/$h$)} & \colhead{(M$\subsun/h$)} 
    & \colhead{Method} & \colhead{Yield $(y)$ }
       }
 \startdata
  25\_256\_0   & 25 & 256   & 98 & $6 \times 10^{7}$ & 0 & - \\
  25\_384\_0   & 25 & 384   & 65 & $2 \times 10^{7}$ & 0 & - \\
  25\_512\_0   & 25 & 512   & 49 & $7 \times 10^{6}$ & 0 & - \\
  25\_768\_0   & 25 & 768   & 33 & $2 \times 10^{6}$ & 0 & - \\
  25\_1024\_0 & 25 & 1024 & 24 & $9 \times 10^{5}$ & 0 & - \\
  25\_256\_1   & 25 & 256   & 98 & $5 \times 10^{7}$ & 1 & 0.025 \\
  25\_384\_1   & 25 & 384   & 65 & $2 \times 10^{7}$ & 1 & 0.025 \\
  25\_512\_1   & 25 & 512   & 49 & $7 \times 10^{6}$ & 1 & 0.025 \\
  25\_768\_1   & 25 & 768   & 33 & $2 \times 10^{6}$ & 1 & 0.025 \\
  25\_768\_1b & 25 & 768   & 33 & $2 \times 10^{6}$ & 1 & 0.005 \\
  25\_256\_2   & 25 & 256   & 98 & $6 \times 10^{7}$ & 2 & 0.025 \\
  25\_384\_2   & 25 & 384   & 65 & $2 \times 10^{7}$ & 2 & 0.025 \\
  25\_512\_2   & 25 & 512   & 49 & $7 \times 10^{6}$ & 2 & 0.025 \\
  25\_768\_2   & 25 & 768   & 33 & $2 \times 10^{6}$ & 2 & 0.025 \\
  50\_512\_0   & 50 & 512   & 98 & $6 \times 10^{7}$ & 0 & - \\
  50\_768\_0   & 50 & 768   & 65 & $2 \times 10^{7}$ & 0 & - \\
  50\_1024\_0 & 50 & 1024 & 49 & $7 \times 10^{6}$ & 0 & - \\
  50\_512\_1   & 50 & 512   & 98 & $6 \times 10^{7}$ & 1 & 0.025 \\
  50\_768\_1   & 50 & 768   & 65 & $2 \times 10^{7}$ & 1 & 0.025 \\
  50\_768\_1b & 50 & 768   & 65 & $2 \times 10^{7}$ & 1 & 0.005 \\
  50\_1024\_1 & 50 & 1024 & 49 & $7 \times 10^{6}$ & 1 & 0.025 \\
  50\_512\_2   & 50 & 512   & 98 & $6 \times 10^{7}$ & 2 & 0.025 \\
  50\_768\_2   & 50 & 768   & 65 & $2 \times 10^{7}$ & 2 & 0.025 \\
  50\_768\_2b & 50 & 768   & 65 & $2 \times 10^{7}$ & 2 & 0.005 \\
  50\_1024\_2 & 50 & 1024 & 49 & $7 \times 10^{6}$ & 2 & 0.025 \\
 \enddata
  \tablecomments{Run: simulation name, given as ($l$)\_($N_{\rm
      cells}^{1/3}$)\_(feedback method), where $l$ is the comoving box
    size and $N_{\rm cells}^{1/3}$ is the number of grid cells along each edge of the
    box.  The total number of grid cells is equal to the number of
    dark matter particles.  $\Delta x$: comoving grid-cell size.
    $m_{dm}$: dark matter particle mass.  Feedback method: 0:
    adiabatic (no radiative cooling, feedback from star formation, or
    ionizing UV background); 1: local method; 2: distributed method.
    Metal yield: fraction of stellar mass returned to gas as metals
    ($m_{\rm metals}/m_{*}$).
  } \label{tab:sims}
\end{deluxetable*}

\subsection{The Enzo Code}

We performed the entire suite of simulations with the Eulerian adaptive
mesh-refinement (AMR), hydrodynamics + N-body code,
\texttt{Enzo}\footnote{http://code.google.com/p/enzo}
\citep{1997WSAMRGMBryan,2004astro.ph..3044O,2005ApJS..160....1O}.  The
equations of hydrodynamics are solved using the ZEUS hydrodynamic solver
of \citet{1992ApJS...80..753S}.  The evolution of
collisionless dark matter is computed with an N-body solver that uses
a particle-mesh (PM) gravity solver
\citep{1985ApJS...57..241E,1988csup.book.....H}.  The AMR
functionality of 
\texttt{Enzo} provides the capability to dynamically add resolution to
regions of the computational domain where it may be required for a
variety of reasons, such as baryonic or dark-matter overdensity,
resolving the Jeans length or shocks.  This makes it possible to
have high resolution in regions of interest while ignoring less
important areas, thus limiting computational costs.  However, since
the WHIM is thought to reside in regions of relatively low overdensity 
($0.1 \lesssim \delta_H \equiv \rho_{b}/\bar{\rho_{b}} \lesssim 100$), a very large fraction 
of the computational domain must be properly resolved.  
Rather than allowing for vast regions of adaptive refinement
throughout the domain, which can introduce significant computational
cost simply to manage to the hierarchy of grid patches, we instead
choose to run with AMR disabled and to perform simulations with large,
static grids, referred to here as ``unigrid'' simulations.

\subsection{Star Formation and Feedback}

Star formation is a process that occurs on physical
scales many orders of magnitude below the resolution limit of
cosmological simulations, yet the products (radiation, thermal energy,
and heavy elements) have significant influence on the evolution of
structure on large scales.  Numerical simulations must employ models
based on our limited understanding of the physical conditions that
produce stars and the impact they have on their surroundings, on the
physical scales relevant to the simulations.  Numerous prescriptions for 
feedback have been introduced with slightly varying parameterizations, (e.g.,
\citet{1992ApJ...399L.113C,2003MNRAS.339..289S,2008MNRAS.383.1210S}),
that are capable of broadly reproducing the observed star formation history
\citep{2006ApJ...651..142H}.

We use a modified version of the prescription of \citet{1992ApJ...399L.113C} in our
simulations, which we describe briefly here.  Any grid cell is capable
of forming stars if the following criteria are met: baryon
overdensity, defined here and only here as $\rho_{b}/(\rho_{c,
    0}(1 + z)^{3}) > 100$, where $\rho_{c, 0}$ is the critical density at
  $z = 0$; 
the velocity divergence is negative; and the
cooling time is less than the dynamical time.  In all other places, we
define baryon overdensity as $\rho_{b}/\bar{\rho_{b}}$.  
The original
prescription of \citet{1992ApJ...399L.113C} 
calls for the additional check of Jeans instability.  However, we find
that in large-scale, fixed-grid simulations, this final criterion is
always met, following satisfaction of the first three requirements.
Therefore, we ignore the check for Jeans instability as a cost-saving
measure.  If all three listed criteria are satisfied, then a ``star particle," 
representing a large collection of stars, is formed within
the grid cell with a total mass,
\begin{equation} \label{eq:m_star}
  m_{*} = f_{*} \ m_{\rm cell} \ \frac{\Delta t}{t_{\rm dyn}} \; , 
\end{equation}
where $f_{*}$ is an efficiency parameter, $m_{\rm cell}$ is the baryon
mass in the cell, $t_{\rm dyn}$ is the dynamical time of the combined
baryon and dark matter fluid, and $\Delta t$ is
the hydrodynamical timestep.  Subsequently, this much gas mass is also
removed from the grid cell as the star particle is formed, ensuring
mass conservation.

While the star particle is formed instantaneously within the
simulation, feedback is assumed to occur over a longer time scale,
which more accurately reflects the gradual process of star
formation.  In a single timestep of the simulation, the amount of mass
from a star particle converted into newly formed stars is given by
\begin{equation} \label{eq:m_star_form}
  \Delta m_{\rm sf} = m_{*} \ \frac{\Delta t}{t_{\rm dyn}} \  \frac{(t - t_{*})}
  {t_{\rm dyn}}  \ e^{-(t - t_{*}) / t_{\rm dyn}}  \; , 
\end{equation}
where $t$ is the current time and $t_{*}$ the formation time of
the star particle.  Stellar feedback is represented by the injection of
thermal energy and the return of gas and metals to the grid, all in
amounts proportional to $\Delta m_{\rm sf}$.  The thermal energy, $e$, and
metal mass, $m_{\rm metals}$, returned to the grid are given by
\begin{equation}
  e = \Delta m_{\rm sf} \ c^{2} \ \epsilon \;,\\
  m_{\rm metals} = \Delta m_{\rm sf} \ y\;,
\end{equation}
where $c$ is the speed of light, $\epsilon$ is the ratio of thermal
energy output to total rest-mass energy of the star particle, and $y$
is the fraction of stellar mass 
returned to the grid as metals.  As per Equation \ref{eq:m_star_form}, 
the rate of star formation, and hence the rate at which thermal energy
and metals are injected into the grid by the star particle, rises
linearly, peaking after one dynamical time, then decays exponentially
after that.  For a given star particle, 30\%, 60\%, and 90\% of the total feedback
production has occurred by 1, 2, and 4 dynamical times, respectively.
For a threshold baryon overdensity (as defined above) of 100,
corresponding roughly to a total overdensity of 600 (assuming the
cosmic ratio of baryonic to dark matter), $t_{dyn} \simeq 1.25\ (1 +
z)^{-3/2}$ Gyr.

We choose parameters for the prescriptions of star formation and feedback
similar to those adopted by \citet{1992ApJ...399L.113C}.  
We assume a star-formation efficiency, $f_{*} = 0.1$, a  25\% fraction of stellar 
mass returned to the grid as gas, and $\epsilon = 10^{-5}$.  Of the
gas returned to the grid, 10\% is 
returned in the form of metals, for a total metal yield, $y = 0.025$, consistent with
the calculations of \citet{1996MNRAS.283.1388M}.  This metal yield, $y = 1/40$, 
is consistent with average values in the Milky Way, with a mean star-formation
rate (SFR) of $\sim3~M_{\odot}$~yr$^{-1}$, a core-collapse supernova rate of
1 SN every 40 years (from pulsar counts), and an IMF-averaged metal yield of 
$\sim3~M_{\odot}$ per supernova.  For a Salpeter IMF, the ratio of
supernova energy to metal ejecta rest-mass energy is $\sim0.0015$
\citep{1980ApJ...237..647B}.  With our choice of $y$ and $\epsilon$,
this implies that approximately 27\% of the total supernova energy is
converted into feedback energy.  As one of the main goals of this
work is to understand the relation between metals and the WHIM, we
have also run three additional simulations with the metal yield 
decreased by a factor of 5 to $y = 0.005$.
\citet{2010MNRAS.402.1536S} investigated the dependence of the global
star formation history on the various star formation and feedback
parameters, finding it difficult to induce a significant deviation.
A parameter study performed as a part of this project showed
the star formation history to be robust to changes in the
prescription parameters.  This effect is primarily due to the fact
that star formation is limited by the amount of cold, collapsing gas,
which is independent of the choice of parameter values.

In this paper, we investigate two different methods for injecting feedback 
into the grid:  (1) depositing all of the feedback  into the grid cell where 
the star particle exists; and (2) distributing an equal amount among the 
central cell and its 26 nearest neighbors.  Since this entire procedure is
designed to account for processes unresolvable by the
simulation, neither method should be considered to be representative
of the actual physical processes that are occurring, but are, rather,
an attempt to mimic the end state of those processes at the multi-kpc
scale.  
Injecting feedback into a single cell has well-known practical
limitations: in particular, the great increase in temperature produced by the injection
of kinetic or thermal energy in a grid cell with reasonably high
density can lead to cooling times that are much shorter than the
hydrodynamical timestep, and thus, over-cooling.  This diminishes the 
ability of the stellar products to propagate away from the source.  This issue 
can be mitigated by making the assumption that feedback is able to spread to 
neighboring grid cells before energy is lost through radiative cooling.  This 
technique was employed by \citet{2006ApJ...650..573C} to mimic the effect of
galactic superwinds.  In this work, we distribute feedback to the neighboring
cells in the form of thermal energy, gas, and metals, and we do not find
it necessary to add kinetic energy.  Hereafter, we refer to
the method of injecting stellar feedback into the central cell and 
its 26 neighbors as the \textit{distributed} method, and the method of
inserting feedback solely into the central cell as the \textit{local} method.

\subsection{Radiative Heating and Cooling}

Of utmost importance to properly modeling the evolution of the WHIM
is an accurate determination of the thermal state of the baryons
within the simulation.  \citet{2009MNRAS.393...99W} showed that
including both the radiative cooling from metals and  the
photoelectric heating from the UV metagalactic background
\citep{2001cghr.confE..64H} is vital for simulations of large-scale
structure.  Unfortunately, a direct calculation of the heating and
cooling rates, including metals, requires solving a network of
rate equations that is far too large to be viable for millions, and
now billions, of grid cells.

We use an efficient computational method, originated by
\citet{2008MNRAS.385.1443S} and similar to \citet{2010MNRAS.tmp.1043S},
of coupling a multidimensional table of
heating and cooling values, precomputed with the photoionization
software, \texttt{Cloudy}\footnote{http://nublado.org} \citep{1998PASP..110..761F},
to a non-equilibrium, primordial chemistry network
\citep{1997NewA....2..181A,1997NewA....2..209A}.  The primordial
chemistry network solves for the evolution of six species (H, H$^{+}$,
He, He$^{+}$, He$^{++}$, and e$^{-}$), including collisional ionization and
photo-excitation/ionization rates, using a backward differencing
formula.  We use \texttt{Cloudy} version 07.02.01 to construct a grid
of heating and cooling rates that vary with density, metallicity,
electron fraction, redshift, and temperature.  The redshift
determines the nature of the UV metagalactic background, given by 
\citet{2001cghr.confE..64H}.  We construct an analogous grid of
values run with H and He and subtract this from the larger grid,
so that only the metal heating and cooling rates remain.  The 
rates are normalized by the H number density times the electron number
density ($n_{\mathrm{H}}\ \times\ n_{\mathrm{e}}$).  
During the simulation, we use the electron density from the primordial
network plus a small, metallicity-dependent correction to calculate
the heating and cooling contributions of the metals.
We turn on the ionizing (UV) background at $z = 7$.  The
primordial chemistry network follows the photoionization of H and He,
and we interpolate from the grid of data made with \texttt{Cloudy} to
include both the heating and cooling due to the metals.  For 
$z > 7$, we use a similar grid of heating and cooling data,
created with no radiation background, assuming collisional ionization
only.  The metal heating and 
cooling terms are included within a coupled chemistry and cooling
solver.  In order to maintain accuracy, the solver subcycles the
cooling and chemistry routines together on
timesteps that are no larger than 10\% of formation/destruction
timescales ($\rho_{i}/\dot{\rho_{i}}$) of e$^{-}$ and H, 10\% of the
cooling time ($e/\dot{e}$, where $e$ and $\dot{e}$ are the internal
energy and its time derivative), and half of the hydrodynamic timestep.
Coupling of the chemistry and cooling solver is particularly crucial 
in regions of high density and metallicity, where the cooling time is
very short and the gas can transition from ionized to neutral within a
single hydrodynamic timestep.

\section{Results} \label{sec:results}

\subsection{The Star Formation History}

\subsubsection{The Effects of Radiative Cooling}

\begin{figure}
  \plotone{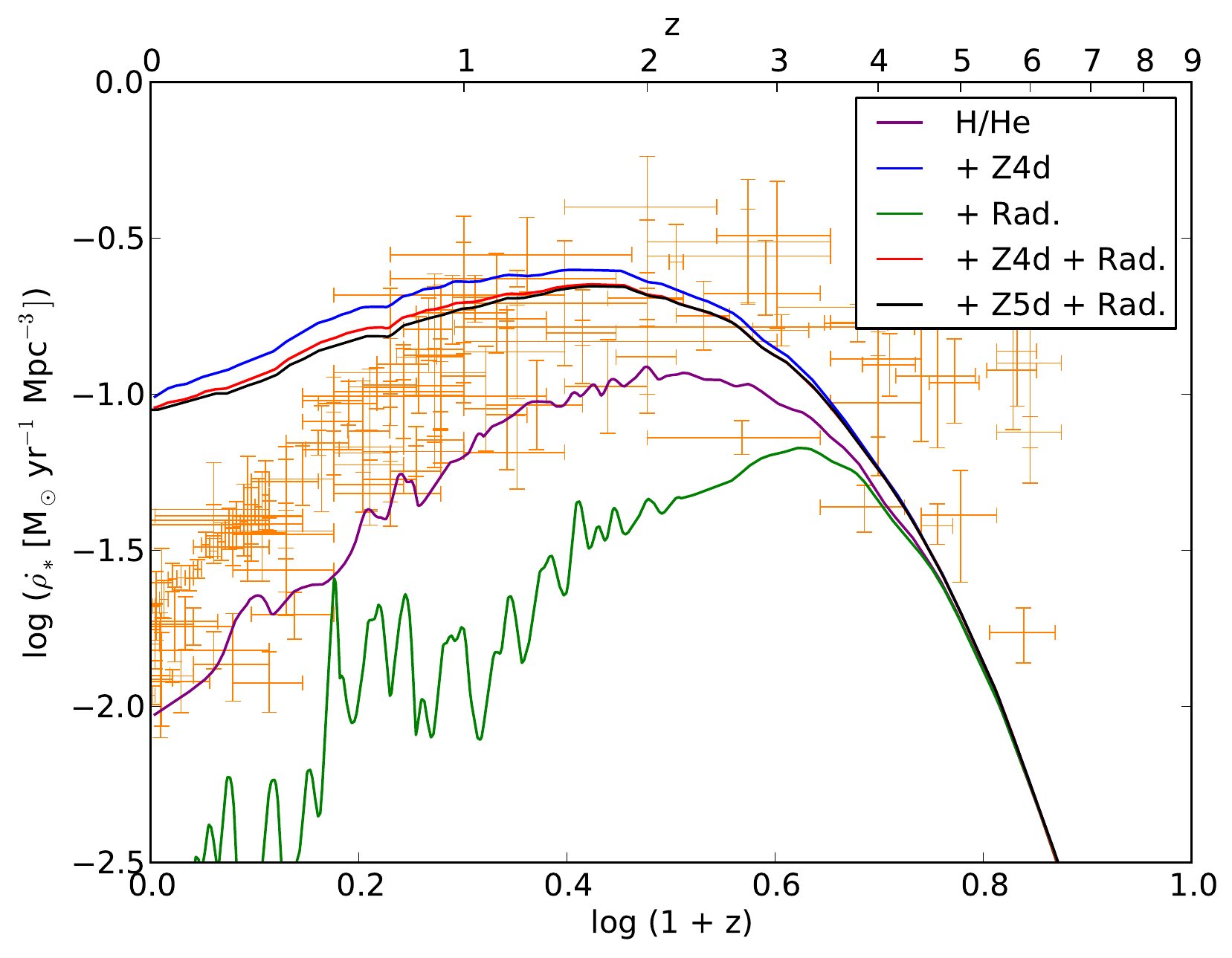}
  \caption{Effect of radiative cooling and radiation background on
    the comoving star formation rate density plotted over the observational data
    compiled by \citet{2006ApJ...651..142H}.  Shown in black is run
    25\_512\_1, with the full cooling model (non-equilibrium
    H/He cooling plus a radiation background and metal cooling that
    includes the effect of the radiation background.  Also shown are
    four additional test simulations with 25 Mpc/$h$ box size and
    512$^{3}$ grid cells with less complete cooling
    methods.  Purple: only the cooling from atomic H/He
    with no radiation background; blue: cooling from H/He
    with no radiation background and metal cooling that does not
    include the effects of the radiation background; green: cooling
    from H/He only plus a radiation background; red: cooling from H,
    He plus radiation background and metal cooling that does not
    account for the radiation background.} \label{fig:sf_cooling_test}
\end{figure}

We first attempt to quantify the effect of the radiative cooling method chosen 
by comparing it with a set of four simulations run with less complete treatments. 
Each of these test simulations has a 25 Mpc/$h$ box with 512$^{3}$ grid cells and 
dark matter particles, with stellar feedback injected into only one cell.  These 
simulations are compared to run 25\_512\_1.  In Figure
\ref{fig:sf_cooling_test}, we plot the comoving 
star formation rate density for run 25\_512\_1 and all of the test
simulations, along with the compilation of observational data on
comoving SFR density by \citet{2006ApJ...651..142H}.  The results of
this experiment can be summarized very simply: a higher cooling rate
results in a higher SFR.  The removal of
radiative cooling from metals results in a significant decrease in the
SFR from $z \sim 5$ to the present.  In the absence of both metal
cooling and the radiation background (purple line in Figure
\ref{fig:sf_cooling_test}), the peak 
in the star formation rate is lower by a factor of a few than the
control run, and occurs slightly earlier (just before $z = 2$, as
opposed to just after.)  The low-redshift slope is also much steeper,
resulting in a SFR at $z = 0$ that is lower by an
order of magnitude.  \citet{2010MNRAS.402.1536S} observe a similar
reduction in the peak SFR, but they do not see the steepening of
the slope at low redshift reported here.

The UV background provides additional heating to
the gas, making it more difficult to form stars.  With only primordial
cooling, the addition of the ionizing background has a far greater influence
on the ability of the gas to form stars than when metals are
included.  In the simulation with the radiation background but no
metal cooling (green line in Figure \ref{fig:sf_cooling_test}), the
SFR is more than half an order of magnitude lower 
than the simulation with no radiation background and no metal
cooling, or over one and a half orders of magnitude below the control
run, which has both a radiation background and metal cooling.  When
metal cooling is included, the presence of the radiation 
background makes much less of a difference, as can be seen with the
blue and black lines in Figure \ref{fig:sf_cooling_test}.  
Photoionization is more effective at reducing the cooling efficiency
of primordial gas because of the relative ease with which H and He
(compared with heavier elements) are fully ionized, thus eliminating
the ability to cool via line emission.  Because primordial coolants
alone are reduced more efficiently by the presence of the radiation
background, it is much more difficult for the gas in these simulations
to continue to cool and form stars.  Finally, if the metal cooling
method does not account for the influence of the 
radiation background, as was neglected in \citet{2009MNRAS.395.1875O}, the
potential exists for a significant overestimate in the total cooling.
This may have been the primary reason that the above work reported the
existence of much colder \OVI\ absorbers than had previously been
claimed or is seen in this work.  The red line in Figure
\ref{fig:sf_cooling_test} represents 
a simulation in which the metal cooling rates do not reflect the
presence of the radiation background.  Interestingly, we do not see
such a dramatic influence on the SFR, as is evidenced by the nearly
identical red and black lines in Figure \ref{fig:sf_cooling_test}.

Not all simulations treat the radiative cooling rate consistently with
the ionizing background and metal transport.  Because the peak of
the radiative cooling rate, at temperatures $4.5 \la \log T \la 6.0$, arises 
primarily from electron-impact excitation of bound states of abundant 
metal ions (\CIII, \CIV, \OVI, \NeVIII),
a strong UV radiation field will photoionize away those coolants and
decrease the cooling.  As can be seen in Figure \ref{fig:sf_cooling_test},
our most complete cooling model over-predicts the SFR at $z < 2$ in this
series of simulations,
owing to the inability of the feedback prescription to drive the metals away 
from their source \citep{2003MNRAS.339..289S}.  As discussed below,
we are able to bring the predicted SFR into better agreement with observations
simply by distributing the thermal energy and metals over multiple grid cells.

\subsubsection{Convergence}

\begin{figure*}
  \plotone{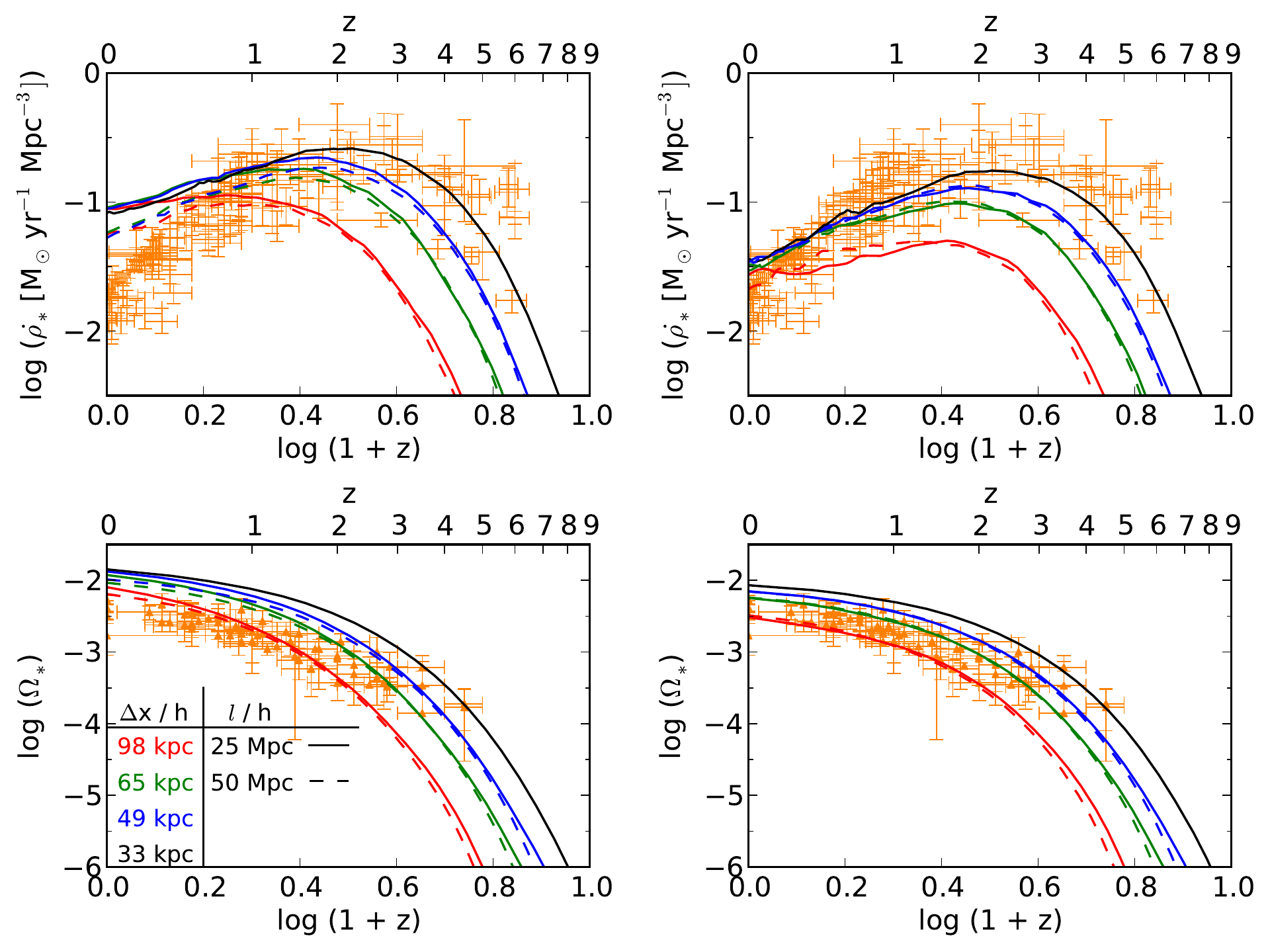}
  \caption{Top panels:  comoving star formation rate (SFR) density versus
    redshift for simulations with varying resolution and box size
    plotted over the observational data compiled by \citet{2006ApJ...651..142H}.
    Simulations with the same resolution are shown with the same 
    color: red, $\Delta x$ = 98 kpc/$h$; green, 64 kpc/$h$; blue, 49
    kpc/$h$; and black, 33 kpc/$h$.  Note that spatial and mass
    resolution are not varied independently in our simulations (see Table \ref{tab:sims}).
    Simulations with 25 Mpc/$h$ box sizes are shown with solid lines, and
    those with 50 Mpc/$h$ box sizes are shown with dashed lines.  
    Bottom panels: stellar density in relation to critical
    density corresponding to top panels plotted over the observational
    compilation of \citet{2008MNRAS.385..687W}.  Left panels denote
    simulations with local feedback, while right panels
    denote those with distributed feedback.} \label{fig:sf_comparison}
\end{figure*}

\citet{2010MNRAS.402.1536S} note the difficulty of matching
the observed SFR at both low and high redshift within
a single simulation.  At high redshift, extreme resolution is required
to resolve the low-mass halos that dominate the global SFR.
At low redshift, a large box size is needed to capture the most
massive objects forming from large-scale perturbations in the
primordial density field that are only recently becoming nonlinear.  
In Figure \ref{fig:sf_comparison}, we plot the SFR versus redshift and
the corresponding stellar density evolution for all 
simulations listed in Table \ref{tab:sims} with metal yield,
$y = 0.025$.  We do not see convergence in the SFR at high
redshift, as noted in previous studies
\citep{2003MNRAS.339..312S,2010MNRAS.402.1536S}.  
Increasing the spatial and gas/dark matter mass resolution, which do not vary
independently in our simulations, allows more low-mass halos to be
resolved at progressively higher redshifts, continually increasing the SFR.
Convergence at $z = 0$ is achieved in our lowest-resolution simulation 
when using the local feedback method.  When using the distributed
method, convergence at $z = 0$ is achieved in our 
second-lowest-resolution simulation.  As the resolution is increased, 
the redshift at which the SFR is converged also increases, as was noted in
\citet{2003MNRAS.339..312S}.  With the local feedback
method, the SFR  of the two highest resolution runs
(25\_512\_1 vs. 25\_768\_1 and 50\_768\_1 vs. 50\_1024\_1) agree to
within 10\% for $z \lesssim 1.5$.  For the distributed-feedback simulations, this
occurs slightly later, at $z \lesssim 1.25$.  Since the interval from 
$z = 1 \rightarrow 0$ represents a majority of the age of the universe,
the resulting stellar densities at $z = 0$ are reasonably converged
for all but the most poorly resolved simulations.

\subsubsection{Feedback and Box Size}

All simulations that employ the local feedback method
overpredict the SFR at low redshift.  This is due to
the over-cooling that occurs when feedback is injected only into the
central cell, unphysically driving up the gas temperature and the
cooling rate \citep{1996ApJS..105...19K,2001MNRAS.326.1228B}.  The
local feedback simulations with equal resolution 
have nearly identical SFRs at high redshift.  However,
those with the larger box size show slightly lower peak 
rates and a steeper decline at low redshift.  This is likely due to
the ability of the large-box simulations to form structures with
higher virial temperatures, in which much of the gas becomes too hot to
form stars.  The simulations that use the distributed method
display the best agreement with observations.  In comparison to
their local feedback counterparts, the distributed-feedback 
simulations have peak SFRs lower by 35--40\%.  

The slope of the SFR with redshift (at $z < 2$) is also noticeably steeper in
the distributed feedback runs.  However, unlike the local feedback
simulations, there are no discernable differences between the small-box
and large-box simulations of equal resolution when using the
distributed method.  Comparing the highest resolution simulations of each
feedback method, we find the use of local method results in the
formation of approximately 1.7 times more stars than the distributed 
method.  In comparison with observational measurements by
\citet{2008MNRAS.385..687W} of the stellar mass density, our simulations still
appear to produce too many stars by a factor of a few.  However,
\citet{2008MNRAS.385..687W} point out that independent observations of
the SFR and the stellar mass density are not consistent with each other; 
the observed  SFR implies a stellar density 2-3 times higher than what is seen.  
The final conclusion is that spreading stellar feedback
over 27 cells, rather than just 1, is largely able to overcome the
classic over-cooling problem, yielding a much better fit to fit to observations.

\subsection{The Halo Mass Function}

\begin{figure*}
  \plotone{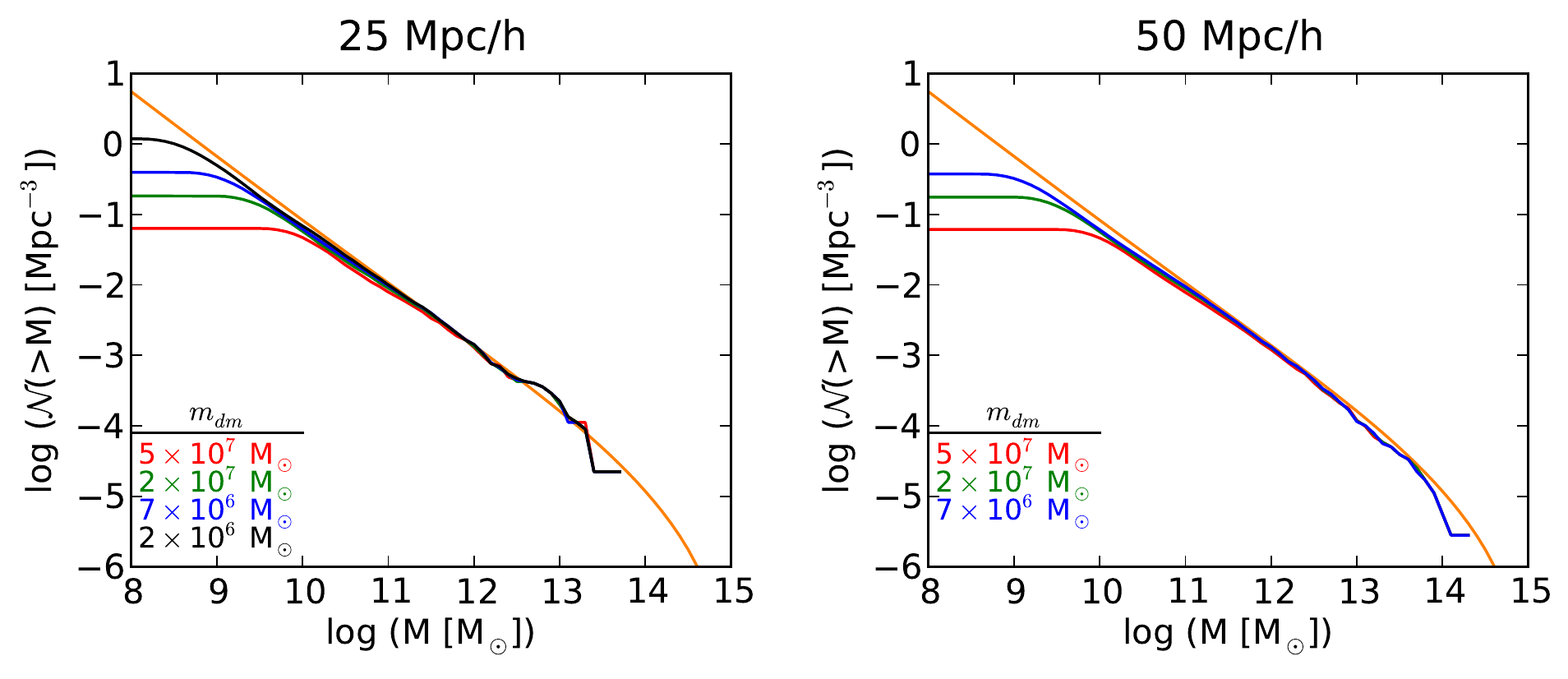}
  \caption{The cumulative halo mass function for all of the simulations with
    distributed feedback and normal metal yield.  Simulations with 25
    Mpc/$h$ boxes are shown in the left panel while simulations with
    50 Mpc/$h$ boxes are shown in the right panel.  Plotted in orange
    is the analytical fit of \citet{2006ApJ...646..881W}.  The colors
    shown here, which indicate the dark matter particle mass within
    the simulation, are consistent with Figure \ref{fig:sf_comparison}.
  } \label{fig:mass_function}
\end{figure*}

In Figure \ref{fig:mass_function}, we plot the cumulative halo mass
function for all the simulations with distributed feedback and normal
metal yield.  To locate halos within our simulations, we use a version
of the HOP halo-finding algorithm \citep{1998ApJ...498..137E},
parallelized by \citet{2010ApJS..191...43S}.  
We show only the mass functions for simulations with distributed
feedback.  For the range of halos captured in our simulations, there
are no notable differences for varying baryonic physics.  In
comparison with the analytical fit of \citet{2006ApJ...646..881W}, we
find our simulations to provide a good match for halos with approximately
400 particles or more, regardless of particle mass.  This result is in good
agreement with numerous, prior \texttt{Enzo} simulations,
e.g., \citet{2005ApJS..160....1O, 2010ApJ...711.1198T, 2010arXiv1006.3559S}.

\subsection{The Evolution of Baryon Phases} \label{sec:baryon_phase}

\subsubsection{Convergence}

\begin{figure*}
  \plotone{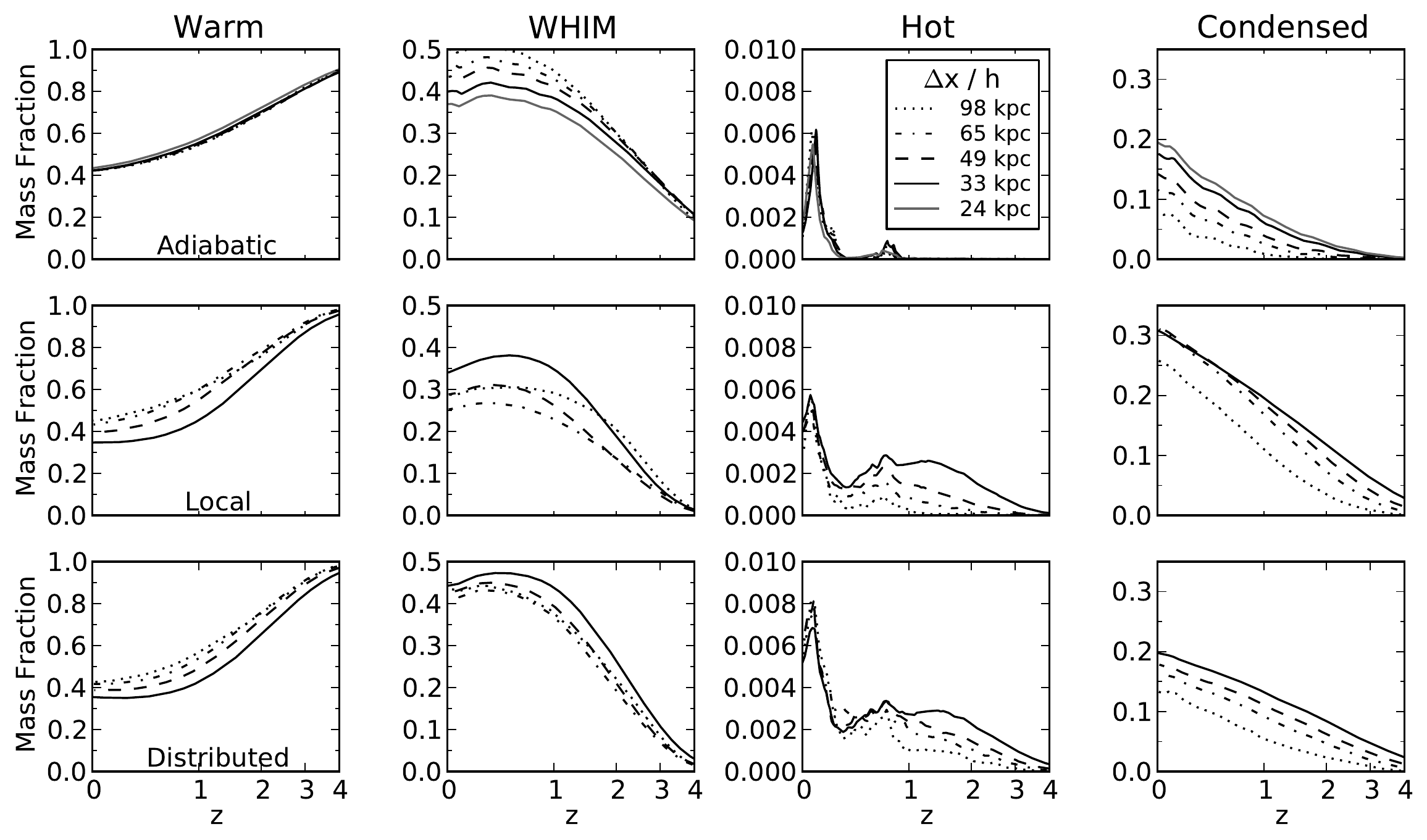}
  \caption{Evolution of baryon mass fraction in the warm ($T < 10^{5}$~K), 
    WHIM ($10^{5}~{\rm K} \le T < 10^{7}$~K), hot ($T \ge 10^{7}$~K), and condensed 
    ($\rho_{b}/\bar{\rho_{b}}$ $\ge$ 1000) phases for 
    the 25 Mpc/$h$ box simulations.  The condensed phase includes both gas with
    baryon overdensity $\delta_H > 1000$ and the total mass in stars.  For the
    warm, WHIM, and hot phases, only gas with $\delta_H < 1000$ is included,
    such that the total is unity.  The top row shows the adiabatic
    simulations, the middle row shows the simulations with local
    feedback, and the bottom row shows the simulations with
    distributed feedback.} \label{fig:mass_fraction_convergence_25}
\end{figure*}

\begin{figure*}
  \plotone{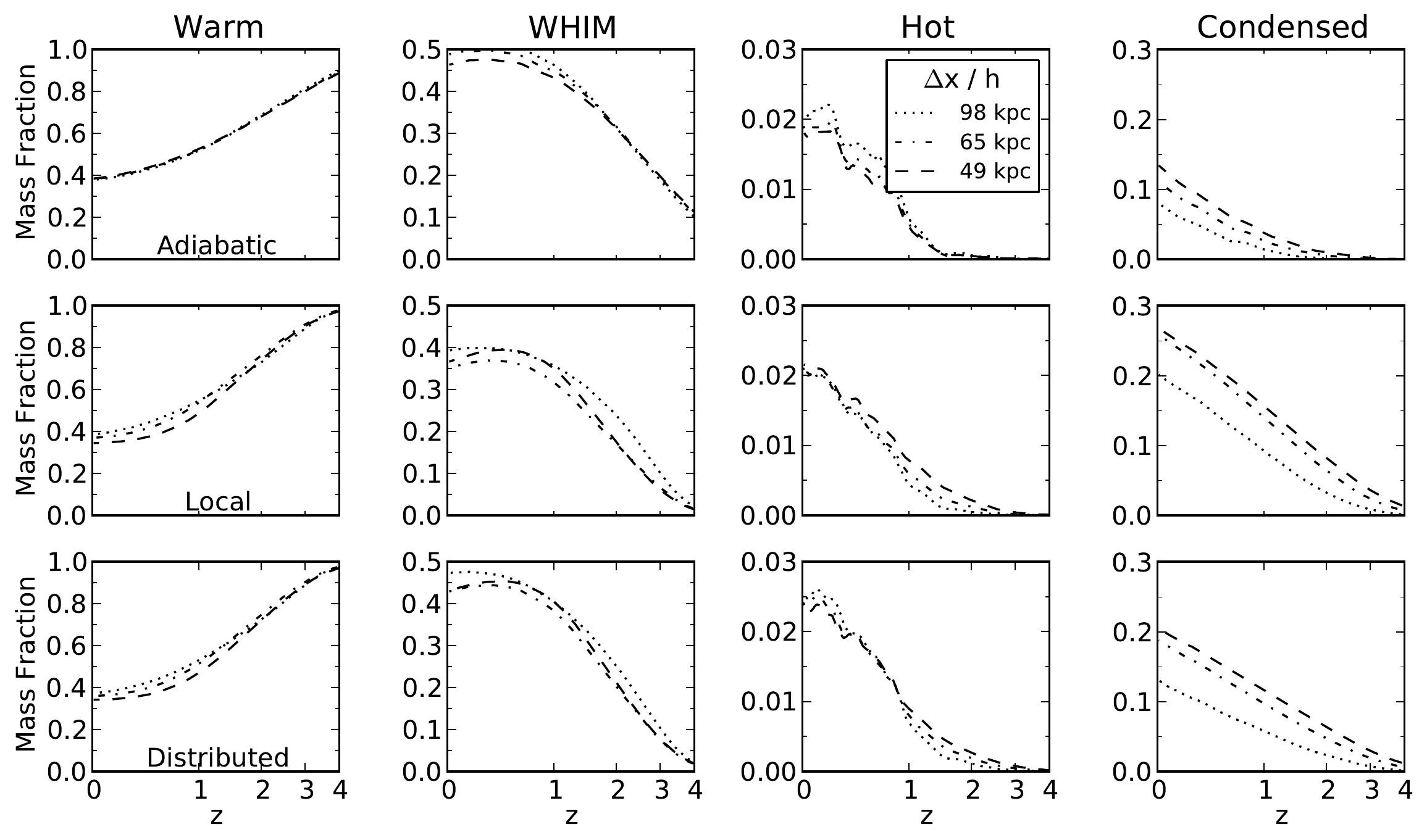}
  \caption{Evolution of baryon fraction in the warm, WHIM,
    hot, and condensed phases for the 50 Mpc/$h$ box
    simulations.  The top row shows the simulations with local
    feedback and the bottom row shows the simulations with distributed
    feedback.  Phase divisions are the same as in Figure
    \ref{fig:mass_fraction_convergence_25}.} \label{fig:mass_fraction_convergence_50}
\end{figure*}

In Figures \ref{fig:mass_fraction_convergence_25} and
\ref{fig:mass_fraction_convergence_50}, we plot the evolution of the baryon
fraction in various IGM phases as a function of redshift for the majority
of our simulations.  We group the plots in rows and columns, showing box size 
and feedback parameter so as to focus on convergence.  We divide the baryons 
into four phases: \textit{warm}
($T \le 10^{5}$~K, $\rho_{b}/\bar{\rho_{b}} <$ 1000); \textit{WHIM} ($10^{5}$~K $< T
\le 10^{7}$~K, $\rho_{b}/\bar{\rho_{b}} <$ 1000); \textit{hot} ($T > 10^{7}$~K,
$\rho_{b}/\bar{\rho_{b}} <$ 1000); and \textit{condensed} ($\rho_{b}/\bar{\rho_{b}} \ge$ 1000).
We also include the total mass in stars in the condensed phase.  In
practice, the gas in the condensed phase is almost entirely warm.  If
we define gas with $\rho_{b}/\bar{\rho_{b}} <$ 1000 as \textit{diffuse}, we see
that  the ratio of diffuse to condensed WHIM gas is generally around 20:1 at $z = 0$.  
The masses of diffuse and condensed hot gas are 
within a factor of a few of each other, but neither contributes
significantly to the total baryon budget within our simulations,
mainly due to the somewhat small box sizes.  Note
that these phase cuts are slightly different from those of other
work, such as \citet{2001ApJ...552..473D} and
\citet{2006ApJ...650..560C}.  However, the results are 
insensitive to these minor differences in definition.  We have also
measured the evolution of the baryon fractions, with the overdensity
threshold dividing the condensed phase from the other three phases set
to 200, and find the variations to be marginal.

In Figure \ref{fig:mass_fraction_convergence_25}, we focus on the
simulations with the 25 Mpc/$h$ box size.  The top row of Figure
\ref{fig:mass_fraction_convergence_25}, which shows the evolution of the
adiabatic simulations, reaffirms the idea that shock heating from
structure formation plays the primary role in the evolution of the WHIM
\citep{1999ApJ...511..521D,2001ApJ...552..473D,1999ApJ...514....1C}.
Without radiative cooling, star formation, and the UV background, the
general trends are similar.  Warm, photoionized gas makes up roughly 90\% of the
total baryon mass at $z = 4$, decreasing to $\sim$40\% by $z = 0$.  This
is offset primarily by an increase in WHIM gas from $\sim$10\% to
40-50\% over the same interval.  In the adiabatic simulations, the
mass fractions of warm and hot gas evolve identically, independent of
resolution.  Since the box size in these simulations is rather small,
massive halos with virial temperatures $T_{\rm vir} \ga 10^{7}$ K are
practically non-existent.  Thus, the fraction of hot gas remains
neglible at all times.  The spike in the hot gas fraction at low
redshift seen in all of the simulations shown in Figure
\ref{fig:mass_fraction_convergence_25} occurs due to a merger.  
One resolution effect that exists in the
adiabatic simulations is a tradeoff between decreasing WHIM gas and
increasing condensed gas with increasing resolution.  As the
resolution increases, the increased gravitational force resolution
allows more low-mass halos to collapse and cross the critical overdensity 
threshold to be considered condensed gas ($\rho_{b}/\bar{\rho_{b}}
> 1000$) before becoming pressure supported.  In addition, higher
resolution allows for smaller-scale perturbations to be resolved in
the initial density field, which will also increase the mass in
collapsed objects.

The addition of the full complement of physical processes to the
simulations significantly alters the convergence properties.  For both
feedback methods, an increase in resolution yields a slightly lower
fraction of warm gas at $z = 0$.  The trend of decreasing WHIM fraction
with resolution seen in the adiabatic runs has reversed.   For $z > 1$,
the fraction of hot gas increases with resolution.  However, by $z = 0$,
the difference in hot-gas fraction is negligible.  As in the adiabatic
simulations, the mass in the condensed phase increases with redshift,
although the precise evolution with redshift is rather different.  In
general, resolution effects produce relatively minor changes in the
evolution of the different phases.  The most severe lack of
convergence exists for the warm and WHIM phases in 
the local feedback simulations.  The percent differences, defined as
$100 \times \left|x_{1} - x_{2}\right| /\left(\onehalf \left(x_{1} +
    x_{2}\right)\right)$, in the warm and WHIM phases at $z = 0$
between the highest and second-highest resolution simulations with 25
Mpc/$h$ box size and local feedback are 12\% and 17\%, respectively.
For the corresponding distributed-feedback simulations, the percent
differences for the warm and WHIM phases are 9.5\% and 4.4\%.  
Thus, it is difficult to argue that the 25 Mpc/$h$ simulations are
completely converged.

The larger box simulations (Figure
\ref{fig:mass_fraction_convergence_50}) show a greater degree of
convergence.  The percent differences for the 50 Mpc/$h$ simulations
analagous to those given above are 6.9\% (warm) and 4.0\% (WHIM) with
local feedback and 5.0\% (warm) and 0.2\% (WHIM) with distributed
feedback.  Since the star formation
rate is the aspect of these simulations that is furthest from
convergence, this may suggest that creation of WHIM gas via
gravitational shock
heating dominates over stellar feedback for larger box sizes.
Unfortunately, due to computational cost, we were unable to
investigate this by running any 
simulations with $l \ge $ 50 Mpc/$h$ at the same resolution as our most
refined 25 Mpc/$h$ simulation.  However, as we continue to optimize
our numerical methods and the available computational resources 
increase, we will perform additional simulations with larger box
sizes and higher dynamic range.

\subsubsection{Redshift Evolution and Feedback} \label{sec:baryon_phase_feedback}

\begin{figure*}
  \plotone{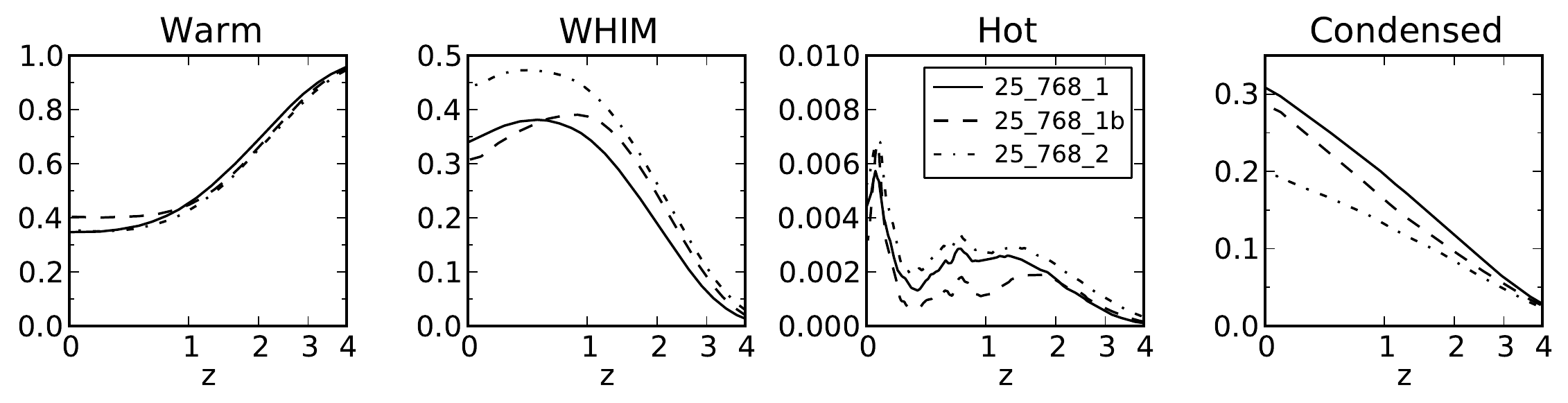}
  \caption{Evolution of  baryon fraction in each phase for the
    three runs with 25 Mpc/$h$ box size and 768$^{3}$ grid cells.
    Shown are two simulations with local feedback, one with metal
    yield, $y = 0.025$ (solid), and one with $y = 0.005$ (dashed), and
    one simulation with distributed feedback and $y = 0.025$
    (dot-dashed).} \label{fig:mass_fraction_25_768}
\end{figure*}

\begin{figure*}
  \plotone{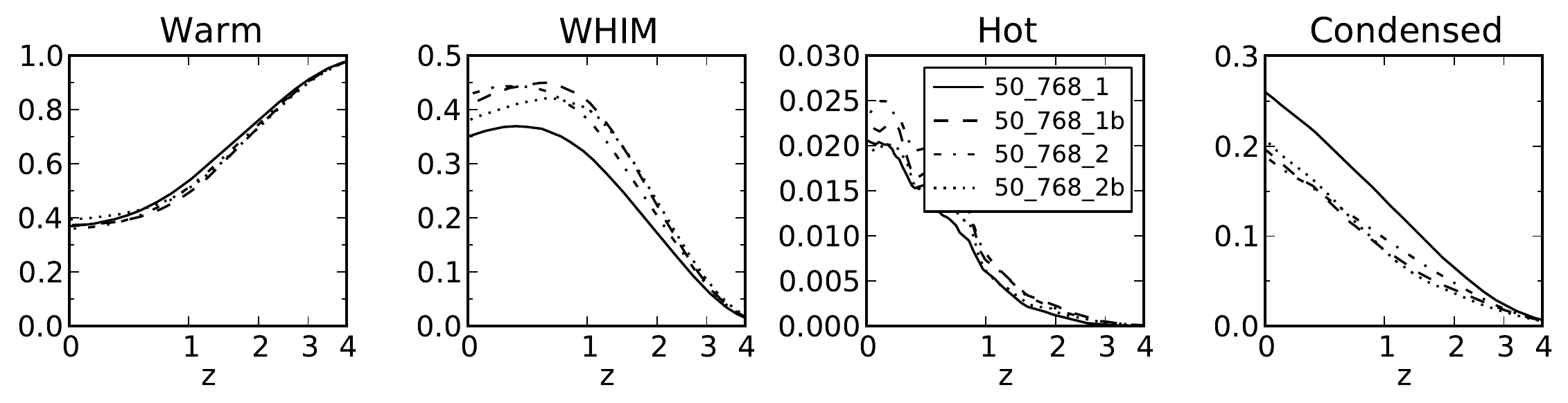}
  \caption{Evolution of  baryon fraction in each phase for the
    four runs with 50 Mpc/$h$ box size and 768$^{3}$ grid cells.
    Shown are two simulations with local feedback, one with $y = 0.025$
    (solid) and one with metal yield $y = 0.005$ (dashed), plus two simulations with
    distributed feedback, one with $y = 0.025$ (dot-dashed) and one
    with $y = 0.005$ (dot).} \label{fig:mass_fraction_50_768}
\end{figure*}

All of our simulations display a behavior, seemingly unique to this work, with respect 
to the evolution of the WHIM gas fraction.  The amount of WHIM peaks at 
$z \approx 0.5$ and is in decline when the simulation reaches the current era
($z = 0$).  This coincides with the fraction of warm gas levelling
out, as opposed to continuing to decrease.  
Although there is no reason to believe that $z = 0$ is a
``privileged epoch," our result is in contrast to most other numerical
studies, which find the WHIM fraction to still be climbing at $z = 0$
\citep{1999ApJ...514....1C, 2006ApJ...650..560C, 2001ApJ...552..473D,
  2010MNRAS.tmp.1043S}.  The recent simulations of
\citet{2010arXiv1005.1451C} show the evolution of the WHIM fraction to 
flatten out at low redshift rather than continuing to increase.  The
``constant-wind'' model of \citet{2010MNRAS.408.2051D} shows a 
decline in WHIM fraction at late time, although it is the only one of
their simulations to display this behavior.  The ``BH'' model of
\citet{2010MNRAS.402.1911T} also shows a decreasing WHIM fraction at
low redshift.  However, that simulation, which was also the only one
in that work to feature this result, was designed specifically to have
feedback from super massive black holes, which we do not include.
Hence, it is somewhat unclear why our results agree.  
Although the adiabatic simulations show this behavior as well, it is more
pronounced in the simulations with cooling and feedback.

\begin{figure*}
  \plotone{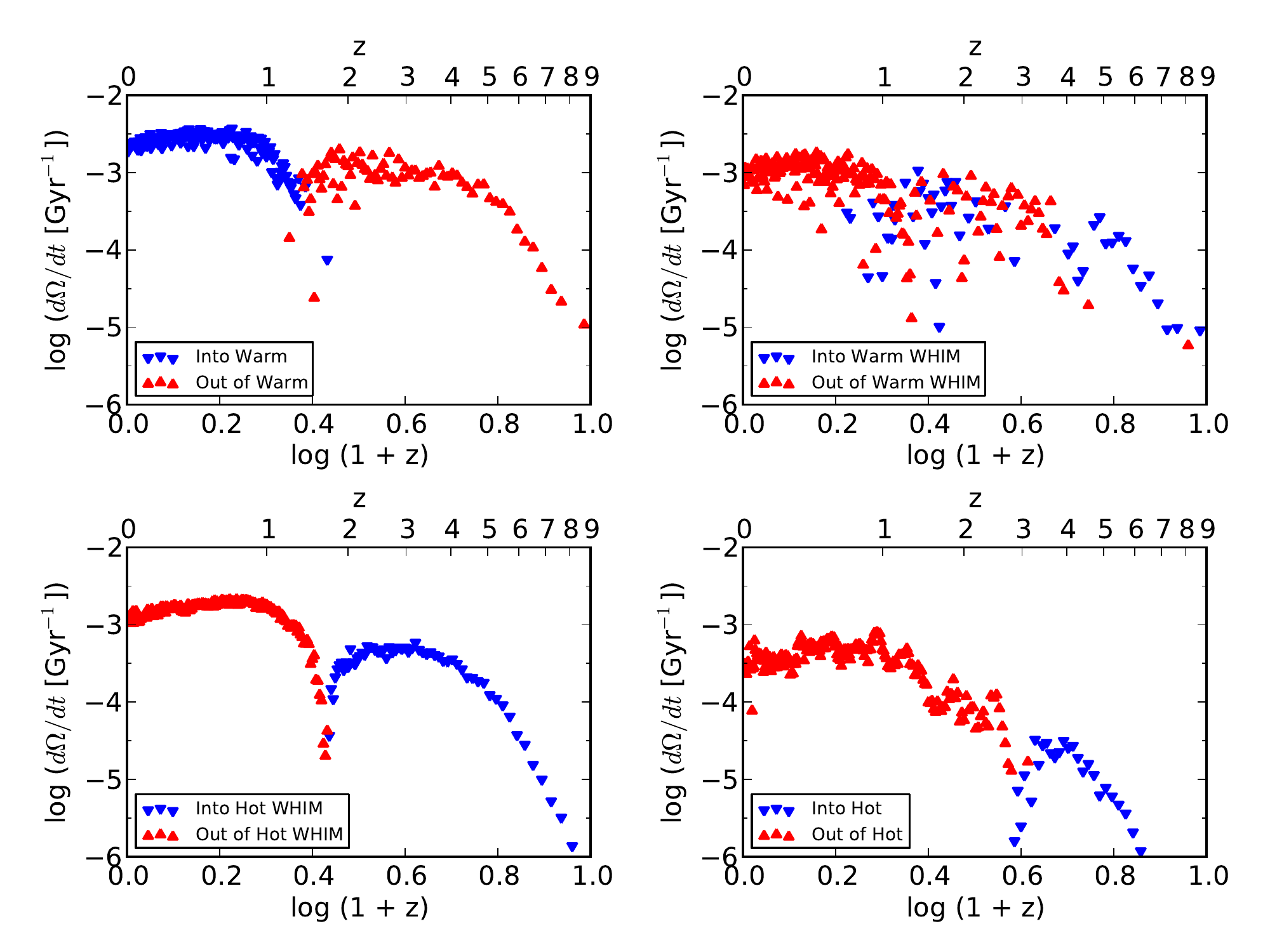}
  \caption{The net flux of baryons into and out of each phase in run
    50\_1024\_2.  Top-left: warm ($T < 10^{5}$ K), top-right:
    warm-WHIM ($10^{5}$ K $\le T < 10^{6}$ K), bottom-left: hot-WHIM
    ($10^{6}$ K $\le T < 10^{7}$ K), bottom-right: hot ($T \ge 10^{7}$
    K).}
  \label{fig:phase_flux}
\end{figure*}

\begin{figure*}
  \plotone{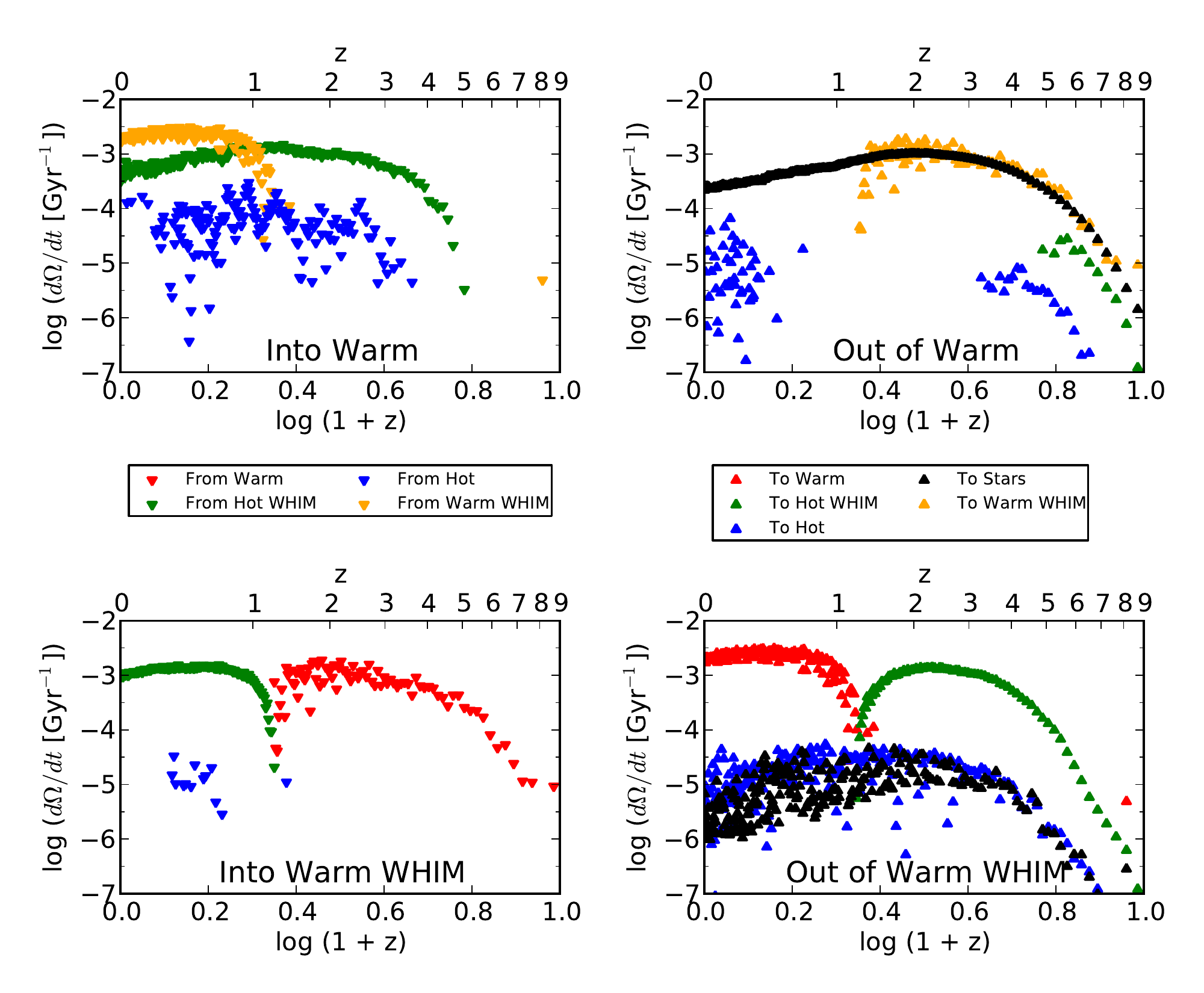}
  \caption{Inputs (left panels) and outputs (right panels) for the
    warm phase (top panels) and warm-WHIM phase (bottom panels) in run
    50\_1024\_2.}
  \label{fig:phase_flux_io}
\end{figure*}

In order to better understand this phenomenon, we attempt to measure
the ``flux'' of gas from one phase to another in a rather crude
manner.  We consider flux between phase A to phase B to exist when a 
single grid cell changes from phase A to phase B when measured in two
temporally nearby data outputs of a simulation.  The value of the flux
from one phase to another is simply taken to be the mass in the later
data output of the cell that has changed phase divided by the time
interval between data outputs.  Unfortunately, we are unable to
account for the change in amount of baryons in a given phase resulting
from a change in density in cells that do not move from one phase to
another.  While this method is clearly flawed, to the extent that the
exact numerical values should not be trusted, it does add useful
qualitative insight.

To disentangle the contributions of cooling and feedback from
cosmological evolution, we first perform this analysis on run
50\_1024\_0, the largest of our adiabatic simuations.  For this
exercise only, we divide the WHIM phase into two phases, the warm-WHIM
($10^{5}$ K $\le T \le 10^{6}$ K) and the hot-WHIM ($10^{6}$ K $\le T
\le 10^{7}$ K).  As the fraction of gas in the warm phase decreases
($z \gtrsim 1$), we measure a net flux \textit{from} the warm phase
\textit{to} the warm-WHIM phase.  At 
the same time, we also see a net flux \textit{from} the warm-WHIM \textit{to} the
hot-WHIM.  At low redshift ($z \lesssim 1$), where the evolution of the warm fraction
appears to flatten out, the flux reverses with warm-WHIM gas now
moving back into the warm phase.  Likewise, hot-WHIM gas now begins
flowing back into the warm-WHIM phase.  In simulations without
radiative cooling, the only means of cooling is adiabatic expansion.
This seems to suggest that, irrespective of cooling and feedback, the
universe undergoes an epoch of heating fueled by shocks from structure
formation followed by an epoch of cooling due simply to the
acceleration of the Hubble flow.  For the cosmological parameters used
in this work, the expansion factor ($E(z) \equiv \left[\Omega_{m} (1 +
  z)^{3} + \Omega_{\Lambda}\right]^{1/2}$) is dominated by
$\Omega_{\Lambda}$ for $z \lesssim 0.4$, which is consistent with the
epoch at which the WHIM fraction begins to decrease in our
simulations.  
When analyzing run 50\_1024\_2 in the same manner, we find a 
similar picture.  In Figure \ref{fig:phase_flux}, we plot the net
phase flux for the warm, warm-WHIM, hot-WHIM, and hot phases for run 
50\_1024\_2.  From this figure, the epochs of heating and cooling are
quite visible, with the reversal of net input to net output occuring
earlier in the higher temperature phases.  In Figure
\ref{fig:phase_flux_io}, we plot the contribution of each phase to the
input and output of the warm and warm-WHIM phases.  It is clear from
Figure \ref{fig:phase_flux_io} that the warm-WHIM is supplied with
material from cooler phases at early times duing the epoch of heating,
and from hotter phases at late times during the epoch of cooling.
Likewise, material leaves the warm-WHIM phase moving mostly into
hotter phases earlier on, and then cooler phases later.

The overall contribution of
radiative cooling and feedback to the evolution of the baryon phases
is less clear, as they add both heating and cooling.  With cooling and
feedback included, the peak in the WHIM fraction occurs slightly
earlier.  In Figures \ref{fig:mass_fraction_25_768} and
\ref{fig:mass_fraction_50_768}, we focus on the influence of feedback
on the evolution of the baryon phases.  For both box sizes, the simulation 
with distributed feedback has roughly 10\% more WHIM  than the simulation with 
local feedback.  The additional WHIM in the distributed feedback simulations 
comes at the expense of condensed gas.  This is another example of the ability of the 
distributed method to overcome the over-cooling problem associated with local
feedback.  Interestingly, the runs with lower metal yield, denoted
with the letter {\it ``b"} in their labels, reach their peak WHIM
fraction at systematically higher redshifts than their higher metal-yield
counterparts.  The WHIM fraction peaks at $z \sim 0.5$ in the
higher-yield runs, but at $z \sim 0.9$ in the lower-yield runs, an
offset of $\sim 2.3$~Gyr.  At higher redshift, the low-yield runs have
slightly more WHIM gas.  However, the WHIM begins to decrease at earlier 
times, and the resulting WHIM fraction at $z = 0$ is lower.  The low-yield runs also
have lower condensed fractions, except in the case of run 50\_768\_2b.  
In every case, these offsets are balanced by a higher fraction of warm
gas.  In comparison to their high-yield analogs, the low-yield runs
reach their peak SFR slightly earlier ($z \sim 2$ as opposed to $z
\sim 1.5$) and have significantly lower SFR at low redshift (roughly
0.5 dex compared to their higher-yield counterparts).  Both of these
can be attributed to the lower radiative cooling efficiency resulting
from the decreased metal yield.  These clues highlight the
importance of accurate treatments of cooling and feedback in the evolution of
the WHIM.  The use of distributed feedback increases the WHIM fraction
by transporting stellar feedback into the IGM with greater
efficiency.  Lowering the metal yield produces a similar effect, but
it also suppresses the SFR more effectively.  Interestingly, the
single simulation of \citet{2010MNRAS.402.1911T} to show a decline in
WHIM fraction at low redshift also happens to be the run in which the
star formation rate is reduced by the greatest amount during that
epoch.  While radiative cooling may
be responsible for some additional amount of gas that is just above 
$10^{5}$ K, cooling into the warm phase, the majority of baryons
residing in the WHIM are at low enough density that their cooling time
exceeds the Hubble time.  Thus, it seems that the role of
radiative cooling is to enhance the ability of dense gas to cool and
continue to form stars, which then inject thermal energy into the IGM,
heating the gas into the WHIM phase.

\subsection{\ion{O}{6} Absorption}

Ultraviolet absorption lines of \OVI\ (1032, 1038 \AA) are believed to be excellent probes
of the cooler portion of the WHIM, as the peak fraction of collisionally ionized
\OVI\ occurs at $T \sim 3\times 10^{5}$ K.  However,
\citet{2009MNRAS.395.1875O} have claimed that the majority of
\OVI\ absorption comes from warm, photoionized gas ($T \sim$ 15,000
K), although their results are likely to be affected by the
significant over-cooling within their simulations.  They suggest that
\OVI\ may not be a reliable WHIM tracer.  
Nonetheless, there exists a number of large 
surveys of \OVI\ absorption lines at low redshift
\citep{2005ApJ...624..555D, 2006ApJ...640..716D,
2008ApJ...679..194D, 2008ApJS..177...39T, 2008ApJS..179...37T,
2008ApJ...683...22T} that provide excellent tests of numerical simulations.
In this section, we make such comparisons.  

\subsubsection{Creating Synthetic QSO Sight Lines} \label{sec:sight_line_creation}

\begin{figure*}
  \plotone{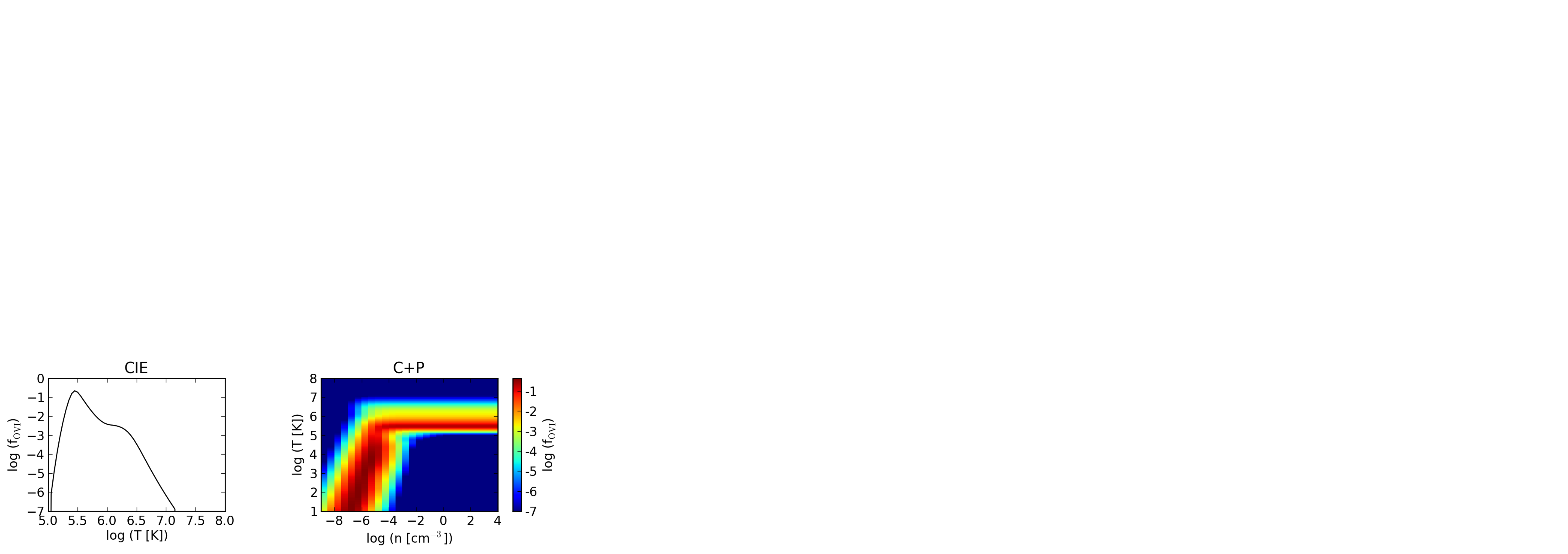}
  \caption{Left:  \OVI\ ionization fraction, $f_{\rm OVI}$, in
    collisional ionization equilibrium (CIE) using data of
    \citet{1993ApJS...88..253S}.  Right:  $f_{\rm OVI}$
    calculated with \texttt{Cloudy} including collisional
    ionization and photoionization from a UV background at $z = 0$
    \citep{2001cghr.confE..64H}.} \label{fig:ion_balance}
\end{figure*}

Our simulations provide a useful database that can be compared to observations
of \OVI\ and other ions along sight lines to active galactic nuclei (AGN).  
We use the \texttt{YT}\footnote{http://yt.enzotools.org}
\citep{SciPyProceedings_46, 2011ApJS..192....9T}
analysis toolkit to construct a set of 500 random AGN sight lines through each of our
simulations.  In order to compare with the \OVI\ absorbers in the survey
of \citet{2008ApJ...679..194D}, we design each sight line to extend
from $z = 0$ to $z = 0.4$, the depth of the survey.  Since the $\delta
z$ corresponding to the length of our simulation boxes is much smaller
than the desired redshift depth, our sight-line generation technique
involves stacking multiple rays cast through simulation boxes from
different epochs.  Before running the simulations, we compute the
exact number of data outputs and their location 
in redshift space required to traverse the comoving distance from 
$z = 0.4$ to $z = 0$, using ray segments no longer than the box
length.  For example, at $z = 0.4$, a comoving distance of 50 Mpc/$h$
corresponds to $\delta z \sim 0.02$.  Thus, in this case, the next
data output in the series is chosen to be at $z = 0.38$.  We continue
this process until reaching $z = 0$.  For the 25 (50) Mpc/$h$ boxes,
47 (23) data outputs are required to span the redshift range.  In this
method, each sight line is able to sample the simulation box at
continuously earlier epochs as it extends further away.  Each
individual subsegment of a given sight line has a random ($x, y, z$)
starting position and ($\theta, \phi$) trajectory so as to minimize
the probability of sampling the same structures more than once per
sight line. 

For each grid cell that is intersected by a ray, we record all
relevant information of the grid cell, such as the baryon density;
temperature; metallicity; redshift; path length, $dl$, of the ray through
the cell; and the corresponding $\delta z$.  Each ray length element, or
lixel, is assigned a negatively incrementing redshift such that the
first lixel has the same redshift as the associated data output and
the last lixel has a redshift exactly $\delta z$ higher than the
redshift of the next data output to be used.  Note that since the ray
segments are randomly oriented and not 
necessarily parallel to any axis of the grid coordinate system, $dl$
is not constant over an entire ray.  The
column density of the $i$'th ionic species of a given element, X, is
then given by $N_{X_{i}} \equiv n_{\rm H} \times (n_{\rm X}/n_{\rm
  H}) \times f_{\rm X_{i}} \times dl$, where $(n_{\rm X}/n_{\rm H})$ is the
elemental abundance of X and $f_{\rm X_{i}}$ is the ion fraction, or
($n_{\rm X_{i}}/n_{X})$.  A coherent structure that may produce a
single absorption line in reality is almost certainly resolved 
over multiple lixels.  In order to minimize cutting structures
into multiple absorbers, we smooth each ray to a constant spectral
resolution ($\lambda/\delta \lambda$) of 5000, which is lower than the
effective spectral resolution of a single grid cell.  Quantities such
as the temperature and metallicity of a smoothed lixel are taken as
the mass-weighted averages of all contributing lixels.  We do not find the
resulting absorbers to be significantly affected by changes of up to
an order of magnitude in either direction to the chosen spectral resolution.

Since we do not track individual heavy
element species, we assume the metals are present in solar abundances
patterns such that $(n_{\rm X}/n_{\rm H}) \equiv Z \times (n_{\rm
  X}/n_{\rm H})_{\odot}$.  For the oxygen abundance, 
we use the value of ($n_{\mathrm{O}} / n_{\mathrm{H}}$)$\subsun$ = $4.9\times 10^{-4}$, as
recommended by \citet{2005ASPC..336...25A}, and we employ two different methods to
calculate the ionization fraction of \OVI.  We first consider the extreme 
case of no ionizing radiation, where \OVI\  is in collisional ionization
equilibrium (CIE) and $f_{\rm OVI}$ is only a function of temperature.  For this,
we use the calculations of \citet{1993ApJS...88..253S}.  We note the
existence of newer calculations performed by \citet{2007ApJS..168..213G}, 
based on updated  ionization and dielectronic recombination rates.
We intend to use these newer rates in our future work.
Fortunately, the values for \OVI\ differ only marginally between
the two studies.  We plot $f_{\rm OVI}(T)$ for CIE in the left panel of Figure
\ref{fig:ion_balance}.  In the second method, we assume the same UV metagalactic 
background \citep{2001cghr.confE..64H} used in the simulations.  We do
not apply any additional renormalization to the intensity of the
spectra.  At $z = 0$, the mean intensity, $J_{\nu, 21}$, of the background in units of
10$^{-21}$ erg cm$^{-2}$ s$^{-1}$ Hz$^{-1}$ is $3.8\times10^{-2}$ at
13.6 eV and $2.6\times10^{-4}$ at 138 eV (the ionization energy of
\OVI).  At $z = 0.4$, $J_{\nu, 21}(13.6\ {\rm eV}) = 0.12$ and $J_{\nu,
  21}(138\ {\rm eV}) = 7.3\times10^{-4}$.  When photoionization is included, the
equilibrium value of $f_{\rm OVI}$ becomes a function of density
as well as temperature.  In the right panel of Figure
\ref{fig:ion_balance}, we show $f_{\rm OVI}(\rho,T)$
calculated with \texttt{Cloudy} for the UV background at $z = 0$.  For
proper number densities $n_H \ga 10^{-2}$~cm$^{-3}$, the
equilibrium value of $f_{\rm OVI}$ is determined solely by electron collisions.  
At lower densities, photoionization begins to dominate, and the metals are typically 
in a higher ionization state at a given temperature than in CIE.
Thus, the temperature of maximum \OVI\ fraction decreases with
decreasing density.  There is significant evolution in the intensity
of the UV background of \citet{2001cghr.confE..64H} 
between $z = 1$ and $z = 0$.  We account for this when constructing the 
synthetic AGN sight lines by calculating $f_{\rm OVI}$ values for each
pixel intersected by the ray segment, using interpolation tables that
vary in density, temperature, and redshift.  Hereafter, we will refer
to this method for calculating $f_{\rm OVI}$ as the collisional +
photoionization (C+P) model. 

\subsubsection{Redshift Distribution of \OVI\ Absorbers}   \label{sec:dndz}

\begin{figure*}
  \plotone{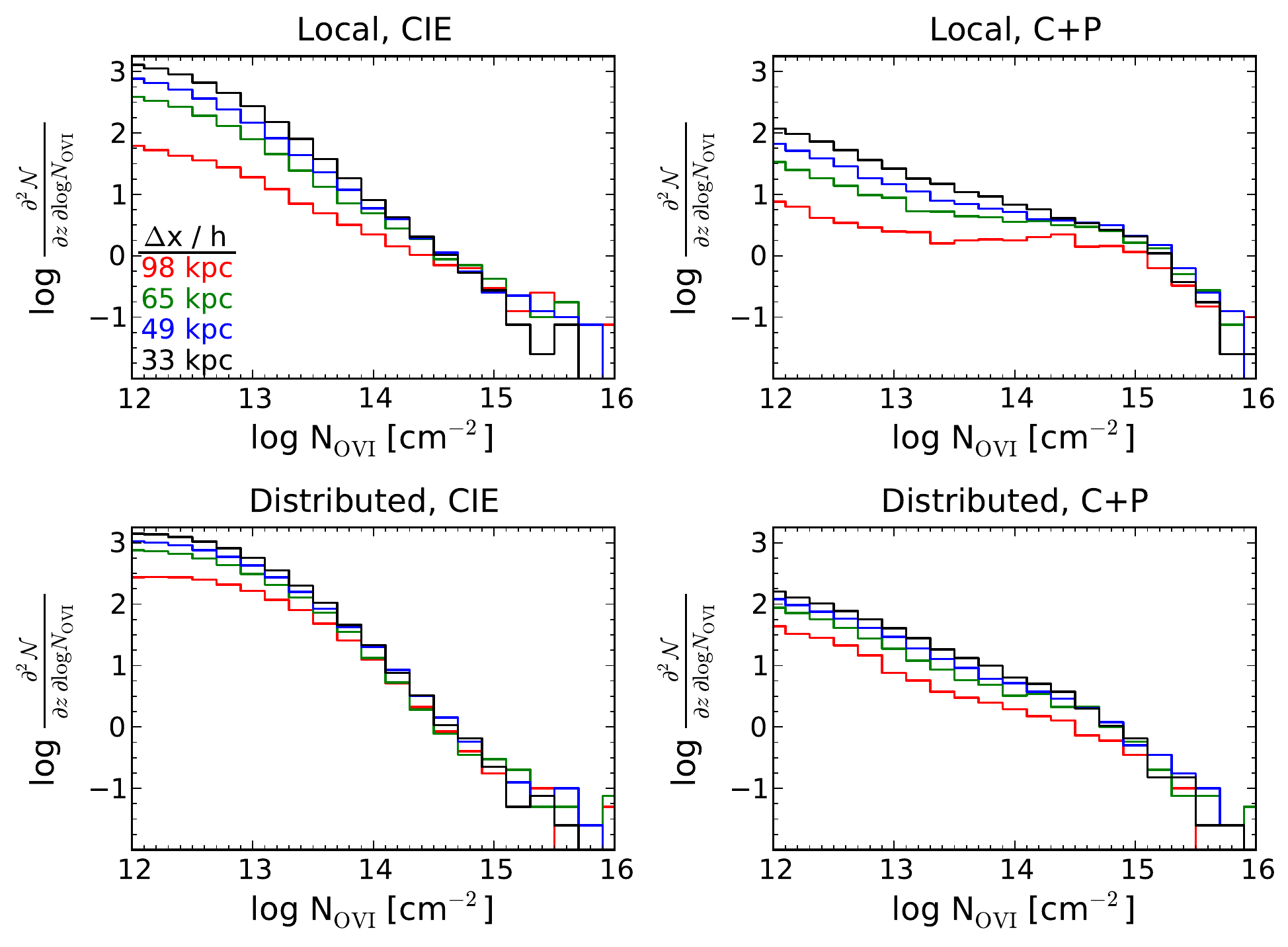}
  \caption{Convergence of $\partial^{2}{\cal N}/\partial
    z \partial(\log$ \NOVI), the number density of \OVI\ absorbers per
    unit redshift and per $\Delta(\log $\NOVI), for the
    simulations with 25 Mpc/$h$ box size and metal yield, $y =
    0.025$.  Line colors denote the resolution of the simulations as
    in Figure \ref{fig:sf_comparison}.  The top two panels show the
    simulations with local feedback, while the bottom panels show
    those with distributed feedback.  The left panels represent \OVI\
    fractions calculated with the CIE model, while the right panels
    represent \OVI\ fractions calculated with the C+P model.} \label{fig:d2ndzdN_OVI}
\end{figure*}

Numerous studies have attempted to recreate the observed number density
per unit redshift of \OVI\ absorbers as a function of column density
to gauge the accuracy of their simulations \citep{2001ApJ...561L..31F,
  2006ApJ...650..573C, 2009MNRAS.395.1875O, 2010arXiv1007.2840T}.  
In Figure \ref{fig:d2ndzdN_OVI}, we plot the distribution of \OVI\
absorbers per unit redshift per log $N_{\rm{OVI}}$ ($\partial^2 {\cal N}
/ \partial z \partial(\log $\NOVI)) for the simulations with 25
Mpc/$h$ box size and metal yield, $y = 0.025$.  We separate runs by
feedback method and plot \OVI\ distributions calculated from both the
CIE and C+P models.  The number of \OVI\ absorbers increases
systematically with increasing resolution.  This is not surprising, as
the star formation rate, and hence the production of metals, also
increases with resolution.  However, it should be noted that
\citet{2010arXiv1007.2840T} show the opposite trend (top-left panel of
Figure B2 in Appendix B of that work), with higher
resolution simulations producing fewer \OVI\ absorbers.  
Just as in the evolution of the baryon phases,
the simulations with distributed feedback are more converged than
those with local feedback.  The number of \OVI\ absorbers at very high
column density ($N_{\rm{OVI}} \gtrsim 10^{15}$ cm$^{-2}$) is consistent at
all resolutions.  This is not surprising, since these absorbers are
probably associated with large, collapsed structures that even the
coarsest simulations are capable of resolving.  The disagreement
between simulations with different resolution is greater for lower
column density.  In all cases, the lowest resolution simulations are
grossly under-resolved with respect to the production of low column
density absorbers, where we do not see
full convergence.  For the three highest resolution
simulations, the increase in the number of \OVI\ absorbers is roughly 
constant with increasing resolution.  It is not completely clear what
resolution is required to reach convergence.  However, as we discuss
below, we are able to achieve an excellent match with the observed
distribution of \OVI\ absorbers with our highest resolution
distributed-feedback simulation using the C+P method, which also
happens to show the greatest convergence.

\begin{figure*}
  \plotone{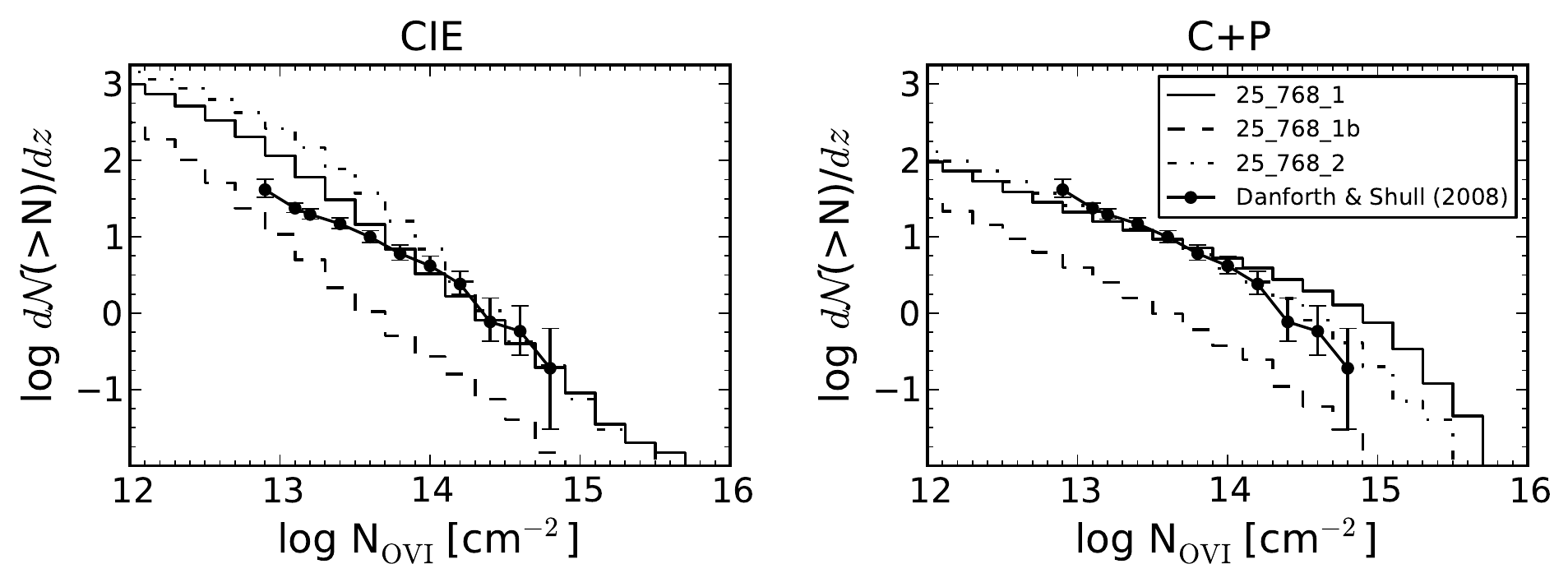}
  \caption{Cumulative number density of \OVI\ absorbers per unit redshift  
   as a function of column density for 500 random sight
    lines extending from $z = 0$ to $z = 0.4$ for runs 25\_768\_1,
    25\_768\_1b, and 25\_768\_2, compared with the observational data of
    \citet{2008ApJ...679..194D}.  Left: $f_{\rm OVI}$ calculated
    assuming collisional ionization equilibrium (CIE).  
    Right: $f_{\rm OVI}$ calculated including both collisional ionization and 
    photoionization (C+P) from a redshift-dependent UV metagalactic background
    \citep{2001cghr.confE..64H}.}
  \label{fig:dndz_OVI_25}
\end{figure*}

\begin{figure*}
  \plotone{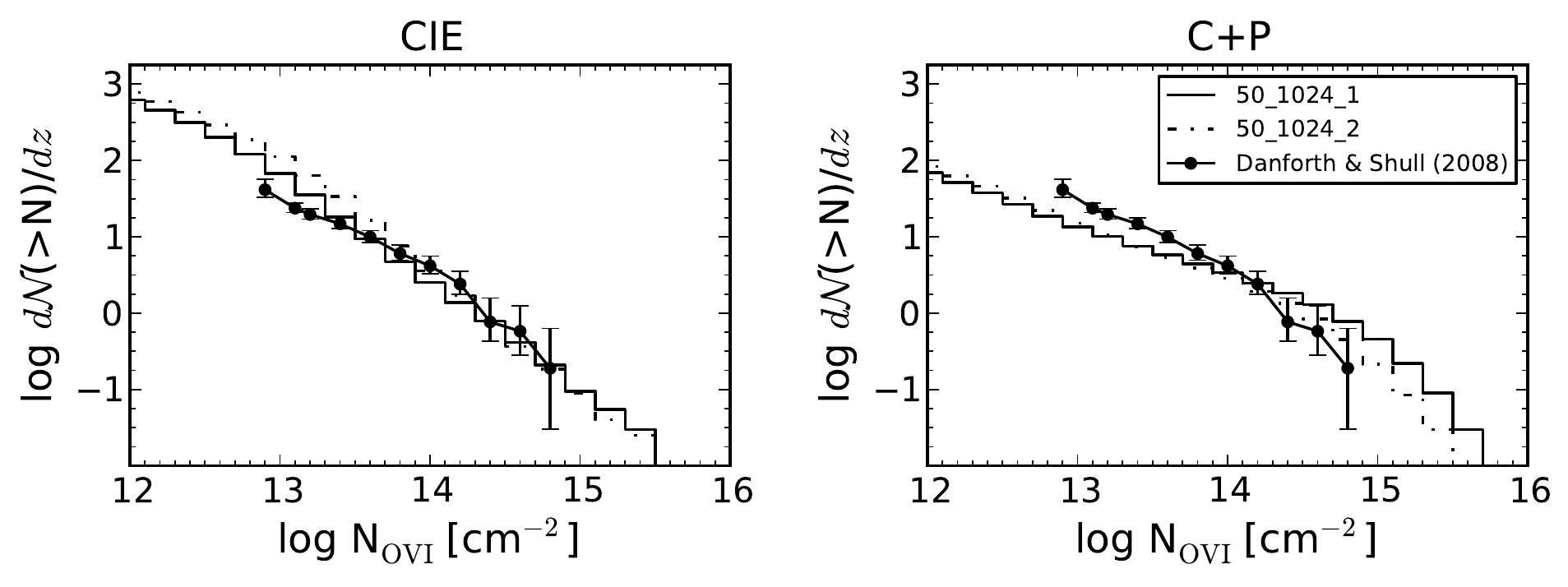}
  \caption{Same as Figure \ref{fig:dndz_OVI_25}, except for runs
    50\_1024\_1 and 50\_1024\_2.}
  \label{fig:dndz_OVI_50}
\end{figure*}

In Figures \ref{fig:dndz_OVI_25} and \ref{fig:dndz_OVI_50}, we plot the cumulative 
number of \OVI\ absorbers, $d{\cal N}(> $\NOVI$)/dz$, given by the integral of 
the bivariate distribution
($\partial^{2}{\cal N}/\partial z \, \partial $\NOVI) over column density \NOVI, 
for our highest-resolution simulations with both 25 and 50 Mpc/$h$ boxes.
For all simulations shown, we plot the $d{\cal N}/dz$ resulting from
$f_{\rm OVI}$ calculations, using both the CIE model (left panels)
and the C+P model (right panels).  We compare our values with
the observations of \citet{2008ApJ...679..194D}.  
The C+P model provides a much better fit to the observed $d{\cal N}/dz$ for
\NOVI\ $ \lesssim 10^{14}\ \mathrm{cm}^{-2}$.  In contrast, the CIE model 
produces an excess of low column density \OVI\ absorbers.  As we discuss 
in \S\ref{sec:ovi_phys}, these absorbers exist primarily in regions of low baryon 
density and are more influenced by ionizing radiation.  For higher column densities
(\NOVI\ $ \gtrsim 10^{14}\ \mathrm{cm}^{-2}$), the C+P model creates slightly too
many absorbers.  This is within the error bars for the simulations
with distributed feedback, but somewhat outside for those with 
local feedback.  The semi-analytical models of
\citet{2005MNRAS.359..295F} show steep drop-off in the number of \OVI\
absorbers at $N_{\rm OVI} \sim {\rm few} \times 10^{14}$ cm$^{-2}$, which is
approximately where we observe $d{\cal N}(> $\NOVI$)/dz$ to drop below
unity nearly universally.  We show in \S\ref{sec:trace_whim} that
the local feedback simulations have significantly more metal 
in regions of high overdensity, which is likely responsible for
the increase in high column density absorbers.  For both feedback
methods, the CIE model is in closer agreement with observations
for the high column density absorbers.  An explanation for this is not
immediately clear and may require the use of full radiation transport 
hydrodynamic simulations instead of the spatially uniform background
employed in this work.

In general, the simulations with the standard metal yield, $y = 0.025$,
display broadly similar results, independent of feedback method.  The
runs using the distributed method show a higher $d{\cal N}/dz$ for
\NOVI\ $ \lesssim 10^{14}\ \mathrm{cm}^{-2}$.  This difference is
more significant when using the CIE model.  The increase in low column
density absorbers from local to distributed feedback models was also
seen in the simulations of \citet{2006ApJ...650..573C}.  
When photoionization is included in the $f_{\rm OVI}$
calculation, the runs with local feedback produce more \ion{O}{6}
absorbers for \NOVI\ $ \gtrsim 10^{14}\ \mathrm{cm}^{-2}$ than the
runs with distributed feedback.  With the CIE model, there is essentially
no difference in the number of high column density absorbers between
the two feedback methods.  The $d{\cal N}/dz$ values for the 50 Mpc/$h$ boxes
show behavior similar to those of the 25 Mpc/$h$ boxes, although they are
slightly lower in general.  This is most likely due to the fact that
the 50 Mpc/$h$ simulations are slightly less converged than the 25 Mpc/$h$
simulations.  However, the $d{\cal N}/dz$ values do not appear to show any
specific dependence on box size.

As expected, lowering the metal yield decreases the number of absorbers at 
all column densities.  However, the decrease in absorber numbers is not 
constant over the range of column density.  
For \NOVI\ $ > 10^{12}\ \mathrm{cm}^{-2}$, both
the CIE and C+P models produce roughly 3 times fewer \OVI\ 
absorbers in the low-yield run.  In the CIE model, this ratio
increases to a peak value of $\sim$12 at \NOVI\ $ \approx
10^{13.3}\ \mathrm{cm}^{-2}$, where it remains constant up to the maximum
column density in the sample.  For the C+P model, this ratio continues
to increase over the entire range of column densities to a value of
$\sim$40 at $N_{\rm max}$.

\subsubsection{$\Omega_{\rm OVI}$} \label{sec:omega_ovi}

\begin{figure*}
  \plotone{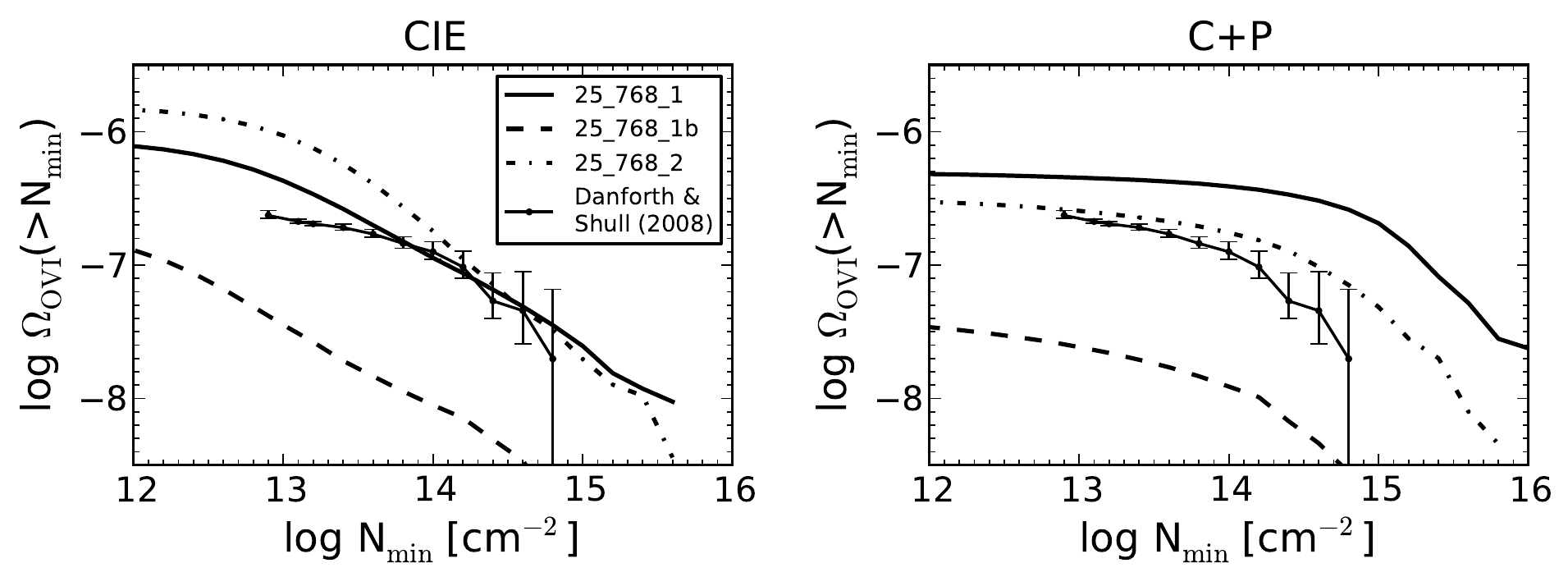}
  \caption{$\Omega_{\rm OVI}$ as a function of minimum \OVI\ 
    column density for 500 random sight
    lines extending from $z = 0$ to $z = 0.4$ for runs 25\_768\_1,
    25\_768\_1b, and 25\_768\_2, compared with the observational data of
    \citet{2008ApJ...679..194D}.  Left: $f_{\rm OVI}$ calculated
    assuming collisional ionization equilibrium (CIE).  
    Right: $f_{\rm OVI}$ calculated including both collisional ionization and 
    photoionization (C+P) from a redshift-dependent UV metagalactic background
    \citep{2001cghr.confE..64H}.}
  \label{fig:omega_OVI_25}
\end{figure*}

\begin{figure*}
  \plotone{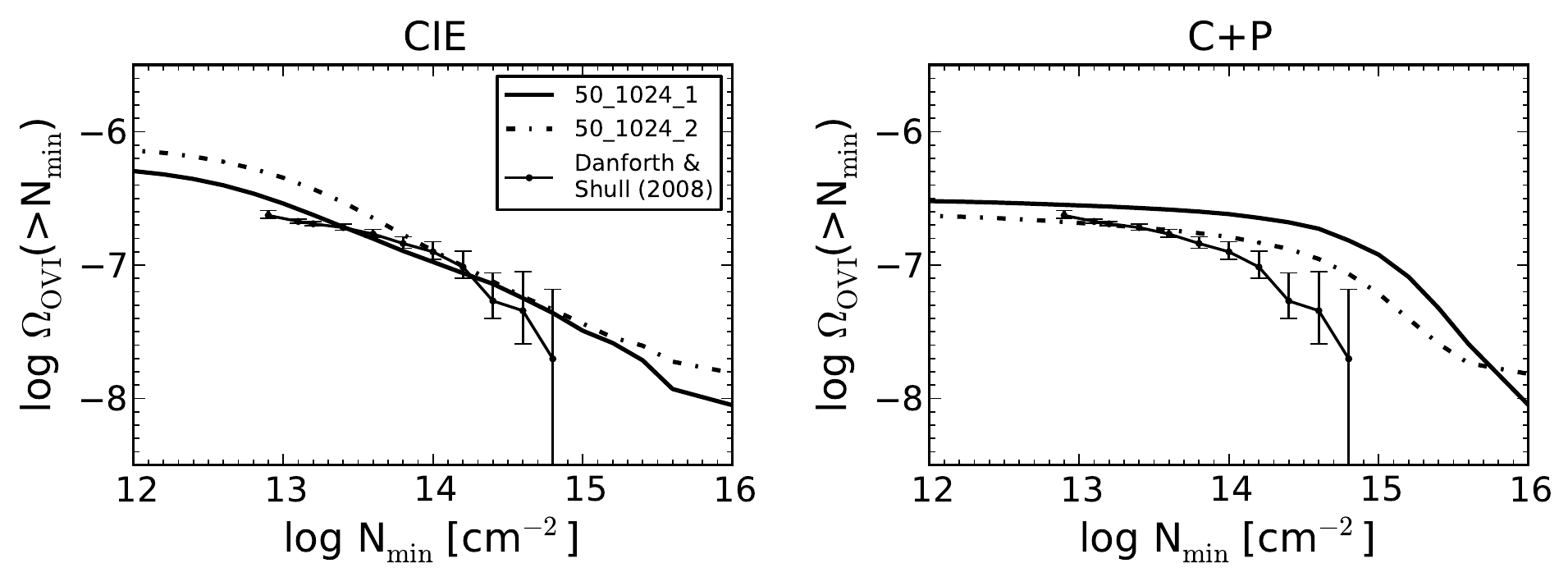}
  \caption{Same as Figure \ref{fig:omega_OVI_25}, except for runs
    50\_1024\_1 and 50\_1024\_2.}
  \label{fig:omega_OVI_50}
\end{figure*}

\begin{deluxetable}{lcccc}
 \tablewidth{0pt}
  \tablecaption{$\Omega_{\rm OVI}(\times 10^{-7})$}
  \tablehead{
    \colhead{Run} & \multicolumn{2}{c}{Simulation Box} &
    \multicolumn{2}{c}{Absorption Lines} \\
    \colhead{} & \multicolumn{2}{c}{at $z = 0$} &
    \multicolumn{2}{c}{log$(N/\rm{cm}^{-2}) >12$ (13)} \\
   \colhead{} & \colhead{CIE} & \colhead{C+P}
    & \colhead{CIE} & \colhead{C+P}
    }
 \startdata
 25\_768\_1   & 17  & 11  & 7.8 (4.3) & 4.8 (4.5) \\
 25\_768\_2   & 20  & 3.5 & 15 (9.4)  & 3.0 (2.5) \\
 50\_1024\_1 & 7.3 & 3.4 & 5.1 (2.9) & 3.0 (2.8) \\
 50\_1024\_2 & 10  & 2.5 & 7.2 (4.5) & 2.3 (2.1) \\
\enddata
  \tablecomments{The contribution of \OVI\ to the closure density of
    the universe for the four highest
    resolution simulations with normal metal yield.  In columns 2 and
    3, $\Omega_{\rm OVI}$ is calculated by summing the total mass in \OVI\
    in the simulation box at $z = 0$ using the CIE (column 2) and C+P
    (column 3) models.  In columns 4 and 5, $\Omega_{\rm OVI}$ is
    calculated from our collection of synthetic \OVI\ absorbers via
    the method described in \S\ref{sec:omega_ovi}.
  } \label{tab:omega_ovi}
\end{deluxetable}

The ratio of the total \OVI\ density to the closure density of the
universe, or $\Omega_{\rm OVI}$, has been calculated for high
\citep{2002ApJ...578...43C, 2004ApJ...606...92S, 2005pgqa.conf..265B}
and low \citep{2000ApJ...534L...1T, 2008ApJ...679..194D} redshift from
the observed density of \OVI\ absorption systems.  Since all of these
surveys have an inherent minimum column density to which they are
sensitive, an important question is how much \OVI\ goes undetected.
Because the column densities of our synthetic \OVI\ absorbers are
calculated directly from the simulation data and not derived from
analysis of spectra, our synthetic survey has essentially infinite
sensitivity, limited only by the spatial resolution of the simulations.

We calculate $\Omega_{\rm OVI}$ from our synthetic \OVI\ survey 
by integrating over the number density of absorbers per cosmological
path length.  This is expressed as
\begin{equation} \label{eqn:omega_ovi}
\Omega_{\rm OVI} = \frac{H_{0}m_{\rm OVI}}{c\rho_{cr,
    0}} \int_{0}^{X_{max}} \int_{N_{min}}^{N_{max}} \frac{\partial^{2}{\cal N}(N)}{\partial
  X\partial \log N} N d\log N dX,
\end{equation}
where $H_{0}$ is the Hubble constant in units of s$^{-1}$, $m_{\rm
  OVI}$ is the \OVI\ mass, $c$ is the speed of light, $\rho_{cr, 0}$
is the critical density of the universe at $z = 0$, and $dX$ is the 
``absorption path length function'', defined by
\citet{1969ApJ...156L...7B} as
\begin{equation} \label{eqn:path_length}
dX \equiv (1 + z)^{2} \left[\Omega_{m}(1 + z)^{3} +
  \Omega_{\Lambda}\right]^{-1/2}\ dz.
\end{equation}
The absorption path length function corrects the number density of
absorbers per unit redshift for the expansion of the universe.  For
the redshift interval from $z = 0$ to 0.4, $X \simeq 0.53$.  Since all
of our synthetic sight lines probe this redshift epoch, the total path
length used in Equation \ref{eqn:omega_ovi} is simply 0.53 multiplied
by the total number of sight lines.  In Figures \ref{fig:omega_OVI_25}
and \ref{fig:omega_OVI_50}, 
we plot $\Omega_{\rm OVI}$ as a function of the minimum \OVI\ column
density.  The values of $\Omega_{\rm OVI}$ are reflective of the
values of \dndz\ and, therefore, the results are very similar.  The
best matches to the observational data of \citet{2008ApJ...679..194D}
come from the simulations with distributed feedback and using the C+P
model for calculating $f_{\rm OVI}$.  Due to the shallow slope of
\dndz\ for low column density absorbers in the C+P model, the
additional \OVI\ mass below current sensitivity limits of $N_{\rm OVI} \sim
10^{13}$ cm$^{-2}$ appears to be negligible.  However,
\citet{2008ApJ...679..194D} find the differential column density
distribution ($\partial^2 {\cal N} / \partial z \partial(\log $\NOVI))
to be proportional to $\sim N^{-2}$, implying equal mass at all column
densities.  Determination of the slope of the \OVI\ distribution at
low column density is a primary science goal for the newly installed
Cosmic Origins Spectrogaph on HST.

We have the luxury of also calculating $\Omega_{OVI}$ via direct
summation of all the grid cells in the simulation box.  This allows us
to determine whether calculation of $\Omega_{OVI}$ in the above manner
produces a biased result.  In Table \ref{tab:omega_ovi}, we list the
values of $\Omega_{OVI}$ calculated through direction summation using
both the CIE and C+P models for each of our four highest resolution
simulations.  We also give the values calculated from our synthetic
absorption line samples for minimum column densities of 10$^{12}$
cm$^{-2}$ and 10$^{13}$ cm$^{-2}$.  The two methods are generally in
good agreement, with direct summation yielding a value roughly
10--40\% higher, with the exception of run 25\_768\_1 in which both
the CIE and C+P summed values are approximately double their counterparts.

\subsubsection{The Physical Conditions of \ion{O}{6} Absorbers} \label{sec:ovi_phys}

\begin{figure}
  \plotone{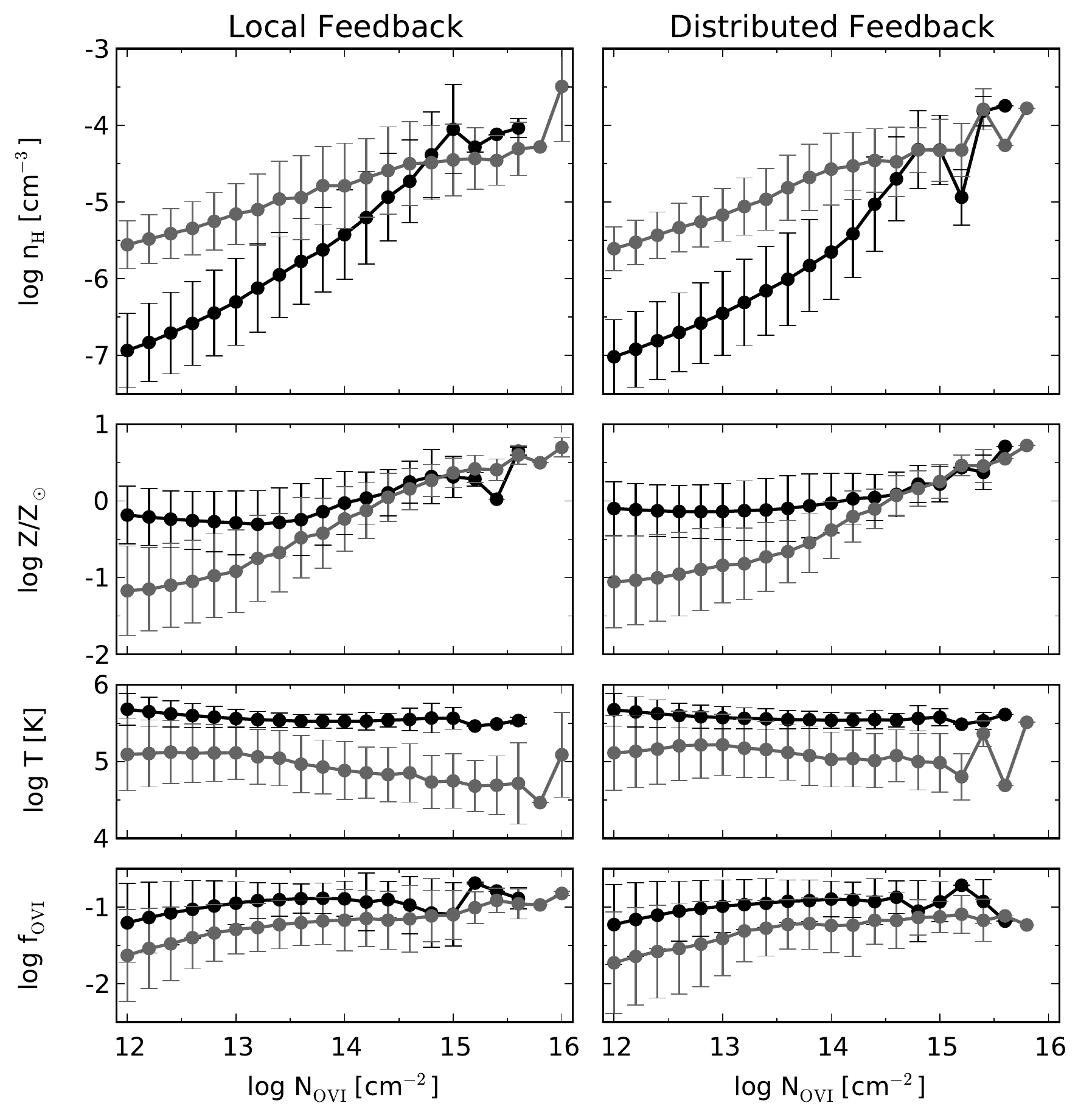}
  \caption{Mean H number density, metallicity, temperature and \OVI\
    fraction at $z = 0$, 
    for two methods of feedback distribution.  Points show
    1~$\sigma$ deviations, with column densities of \OVI\ absorbers shown
    in Figure \ref{fig:dndz_OVI_25} (25 Mpc/$h$ boxes with metal yield $y = 0.025$).
    Black lines show statistics for \ion{O}{6} fractions using 
    CIE model and grey lines represent \ion{O}{6} fractions calculated with
    C+P model.}
 \label{fig:absorber_stats_25}
\end{figure}

\begin{figure}
  \plotone{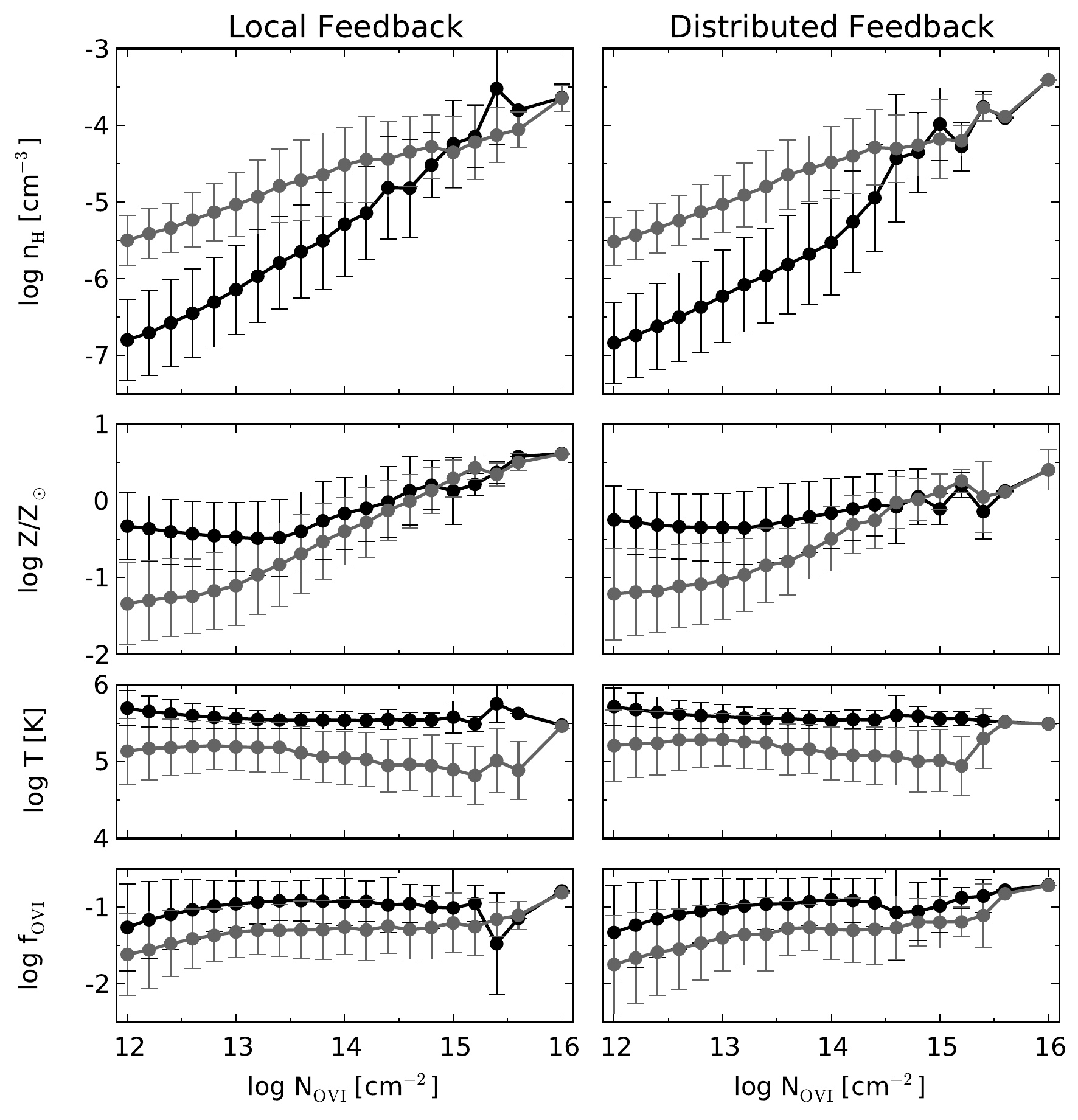}
  \caption{Same as Figure \ref{fig:absorber_stats_25}, but for the
    runs with the 50 Mpc/$h$ boxes shown in Figure \ref{fig:dndz_OVI_50}.}
  \label{fig:absorber_stats_50}
\end{figure}

In Figures \ref{fig:absorber_stats_25} and
\ref{fig:absorber_stats_50}, we examine the physical conditions of the
\ion{O}{6} absorbers plotted in Figures \ref{fig:dndz_OVI_25} and
\ref{fig:dndz_OVI_50}.  The choice of $f_{\rm OVI}$ ionization model has
considerable influence on the average physical conditions of the 
\ion{O}{6} absorbers as a function of their column density.  For
both the CIE and C+P models, the average gas density per absorber rises
with column density.  For the CIE model, we find $n_{H} \propto
N_{\rm OVI}^{0.9}$, and for the C+P model, we find $n_{H} \propto
N_{\rm OVI}^{0.5}$.  The slope of this increase is significantly
steeper with the CIE model.  At the minimum column density measured,
the average gas density associated with \OVI\ absorbers from the CIE
model is approximately 1.5 orders of magnitude lower than with the C+P
model.  For \NOVI\ $ > 10^{15}\ \mathrm{cm}^{-2}$, the associated gas density 
is roughly similar for both models,  with $n_{H} \approx 4.5 \times 10^{-5}$ cm$^{-3}$ or
$\delta_H \approx 240$.  With WMAP-7 parameters,
$\Omega_b h^2 = 0.02260 \pm 0.00053$, the baryon density $\rho_b$ and
corresponding hydrogen number density $n_H$ can be written
\begin{eqnarray}
   \rho_b  &=& (4.26 \times 10^{-31}~{\rm g\ cm}^{-3}) (1+z)^3 \, \delta_H  \nonumber \\
   n_H      &=& (1.90 \times 10^{-7}~{\rm cm}^{-3}) (1+z)^3 \, \delta_H  \; , 
\end{eqnarray}
where $\delta_H$ is the overdensity factor.

\begin{figure*}
  \plotone{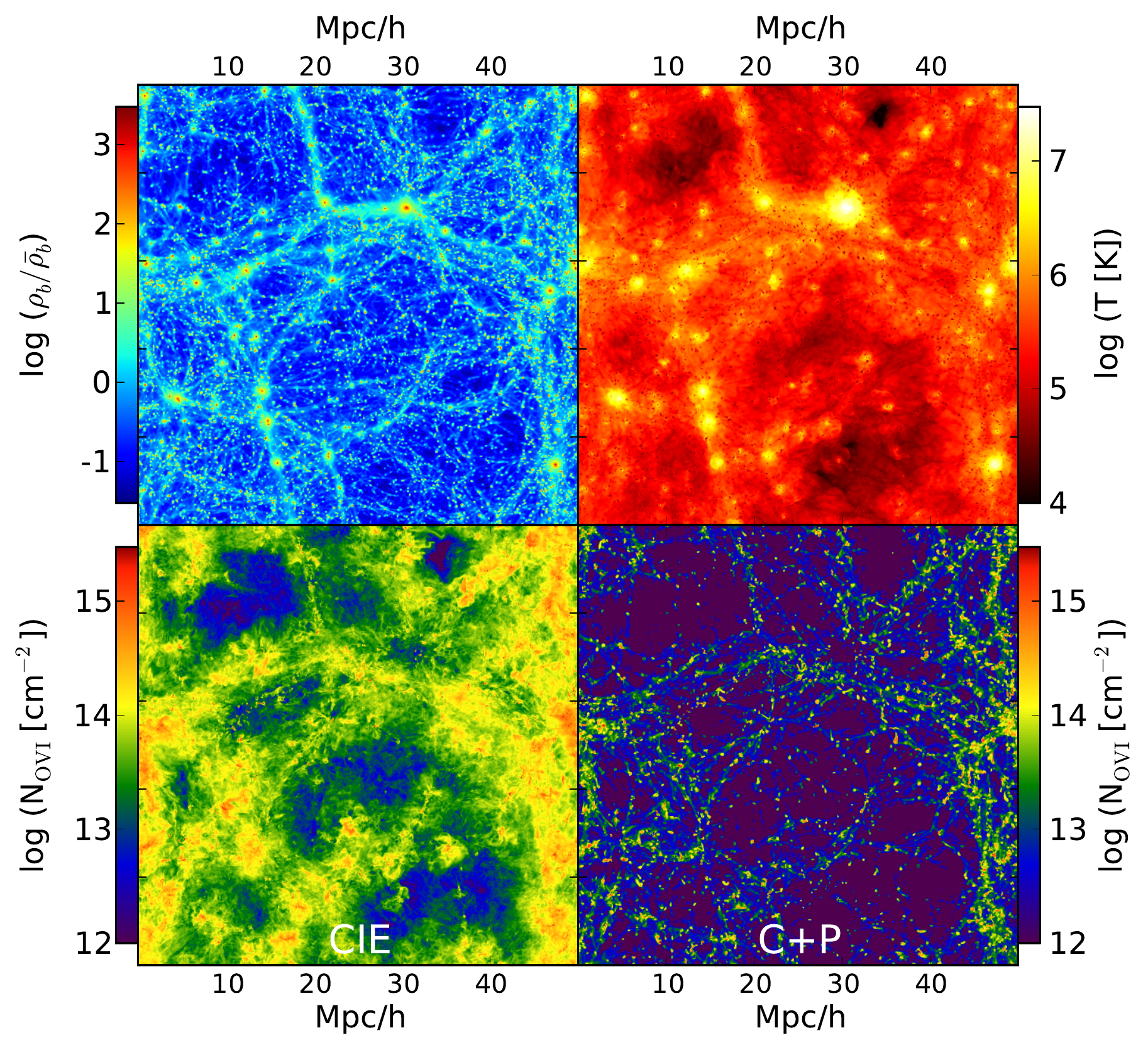}
  \caption{Projections through the full simulation box of run
    50\_1024\_2 at $z = 0$ of mass-weighted mean baryon overdensity
    (top-left), mass-weighted mean temperature (top-right), and
    \ion{O}{6} number density, with $f_{\rm OVI}$ calculated assuming
    CIE (bottom-left) and including photoionization (C+P,
    bottom-right).  Projections are constructed by summing the values
    in all grid cells along the line of sight.  For weighted
    projections, each pixel in the image is calculated as
    $\sum\limits_{i} m_{i} w_{i} / \sum\limits_{i} w_{i}$, where
    $m_{i}$ is the projected quantity and $w_{i}$ is the weighting
    quantity.  For more information, see \citet{2011ApJS..192....9T}.}
  \label{fig:projections_50_2}
\end{figure*}

The spatial difference between \OVI\  in the CIE model and the C+P
model is made clear in Figure \ref{fig:projections_50_2}, where we
show projections through the full simulation box at $z = 0$ of
mass-weighted mean baryon density and temperature, 
along with \OVI\ column density, 
using both $f_{\rm OVI}$ models for run 50\_1024\_2.  With the CIE
model, \OVI\  exists spatially much further from collapsed structures, 
where the gas temperature is in the optimal range for \OVI.  However,
with the C+P model, gas that 
would have a high $f_{\rm OVI}$ in CIE is instead
photoionized to higher ionization states.  In this case, \OVI\ 
is confined primarily to IGM filaments.

With the CIE model, the metallicity of \OVI\ absorbers is relatively independent of 
column density and consistent with $Z = Z\subsun$ up to 
\NOVI\ $ \approx 10^{15}\ \mathrm{cm}^{-2}$.  With
the C+P model the metallicity rises with column density from
$\sim10^{-1}$ $Z\subsun$ at \NOVI\ $ = 10^{12}\ \mathrm{cm}^{-2}$ to $Z\subsun$
at 10$^{15}\ \mathrm{cm}^{-2}$.  The C+P model results are similar to the
findings of \citet{2006ApJ...650..573C}, who see a rise from just
over 0.1 $Z\subsun$ to slightly under $Z\subsun$.  
The density and metallicity values from the C+P models are broadly
consistent with the results of \citet{2009MNRAS.395.1875O}, who employ
a similar method for calculating $f_{\rm OVI}$.  The average absorber
temperature remains fairly constant as a function of column for both
$f_{\rm OVI}$ models.  For the CIE model, the average temperature is
$\sim 3 \times 10^{5}$ K, with a small variance.  This is not
surprising, as the peak value of $f_{\rm OVI}$ in CIE is sharply peaked around
this temperature.  For the C+P model, the average absorber temperature
is $\sim 10^{5}$ K.  Unlike the density and metallicity values, this
is in considerably less agreement with \citet{2009MNRAS.395.1875O},
who find the mean absorber temperature to be $\sim15,000$ K.  This is
most likely due to the higher cooling rates in that work resulting
from not accounting for the radiation background when computing the 
cooling from metals.  The mean 
value of $f_{\rm OVI}$ is roughly 0.1 with the CIE model over the entire
range of column densities.  The values of $f_{\rm OVI}$ from the C+P
model are typically lower than the CIE values by $\sim 0.5$ dex for 
\NOVI\ $ \lesssim 10^{15}\ \mathrm{cm}^{-2}$.  The highest column density
absorbers (\NOVI\ $ \gtrsim 10^{15.5}\ \mathrm{cm}^{-2}$) tend to exist as a
result of particularly optimal conditions for \OVI: high gas densities in
excess of 10$^{-4}\ \mathrm{cm}^{-3}$, super-solar
metallicities, and values of $f_{\rm OVI}$  near the maximum
allowable in either CIE or C+P models.

\begin{figure*}
  \plottwo{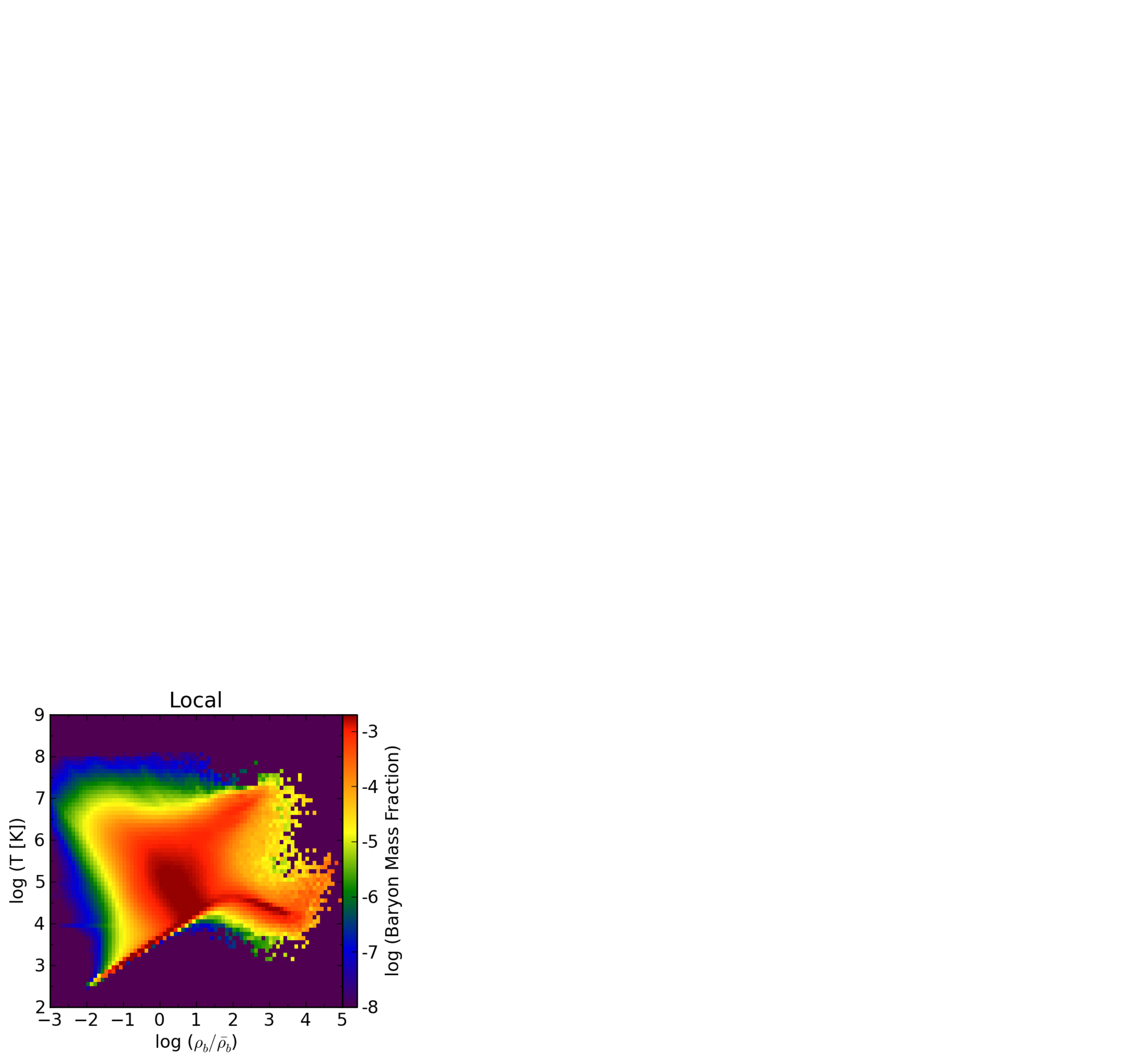}{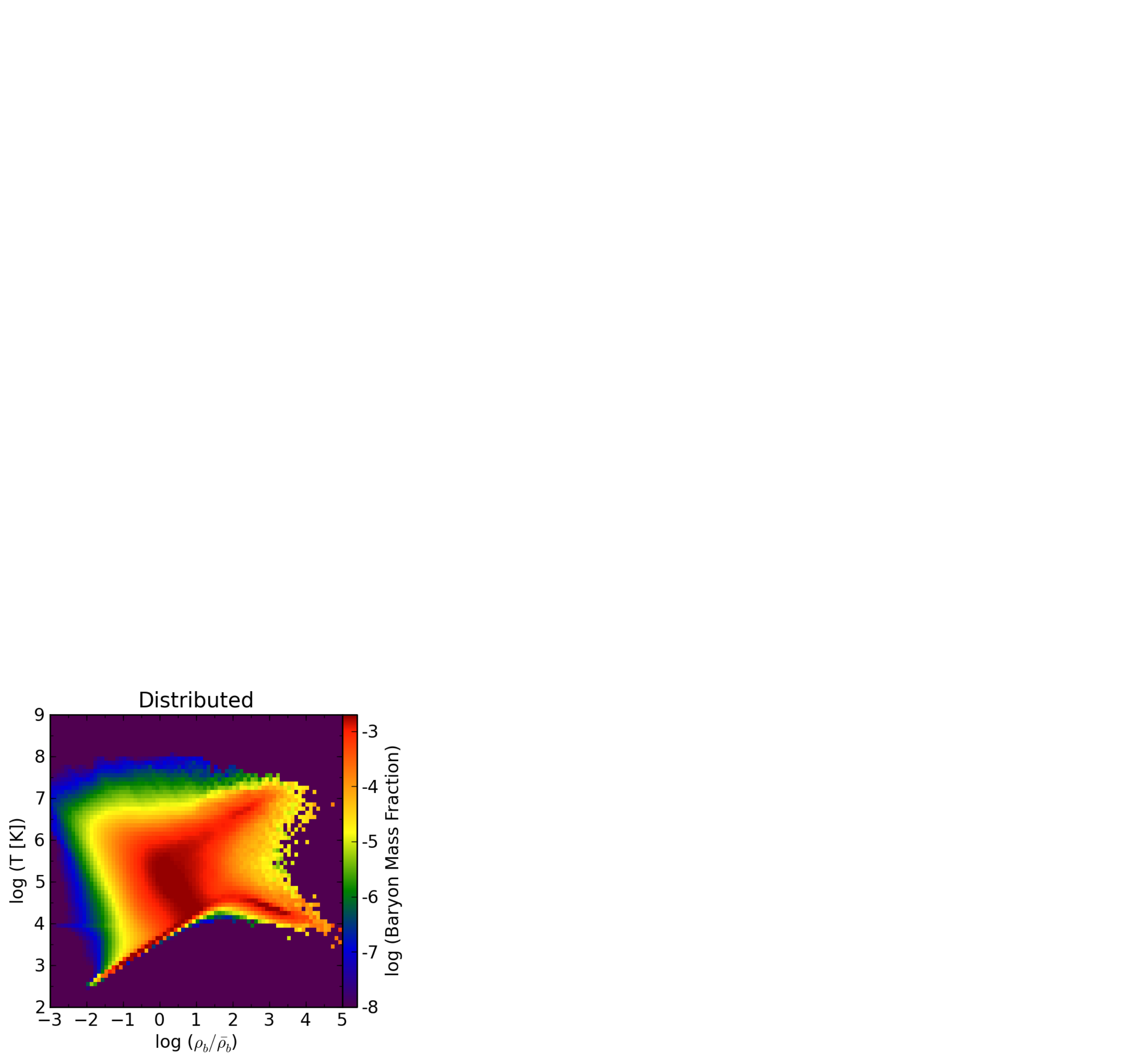}
  \caption{Fraction of total gas-phase baryon mass in two-dimensional
    bins of overdensity ($\rho_{b}/\bar{\rho_{b}}$) and temperature
    for runs 25\_768\_1 (left) and 25\_768\_2 (right) at $z = 0$.  The
    concentration of baryons in the thin red line extending from
    $\rho_{b}/\bar{\rho_{b}} \sim 10^{-2}$ and $T \sim 300$ K up to
    $\rho_{b}/\bar{\rho_{b}} \sim 10^{2}$ and $T \sim 10^{4.5}$ K
    represents unenriched (see Figure \ref{fig:T_OD_Metals_25} for an
    indication of metallicity) IGM lying on an adiabat.  Radiative cooling
    becomes important at higher densities, as evidenced by the
    declining baryon temperature.  The baryon concentration in the
    regime of $10^{2} \lesssim \rho_{b}/\bar{\rho_{b}} \lesssim
    10^{4}$ and $10^{4}$ K $\lesssim T \lesssim$ $10^{5}$ K depicts
    collapsing halos, galaxies, and star-forming regions.
    Virialization shocks and stellar feedback heat gas up to $\sim
    10^{7}$ K where it travels out into the IGM, cooling
    adiabatically.  The adiabat extending from
    $\rho_{b}/\bar{\rho_{b}} \sim 10^{3}$ and $T \sim 10^{7}$ down
    into the broad plume of baryons between $10^{5}$ K and
    $10^{7}$ K is the WHIM.  The fraction of total baryon mass in gas
    (not stars) is 75\% for run 25\_768\_1 and 85\% for run 25\_768\_2.}
  \label{fig:T_OD_Baryons_25}
\end{figure*}

\begin{figure*}
  \plottwo{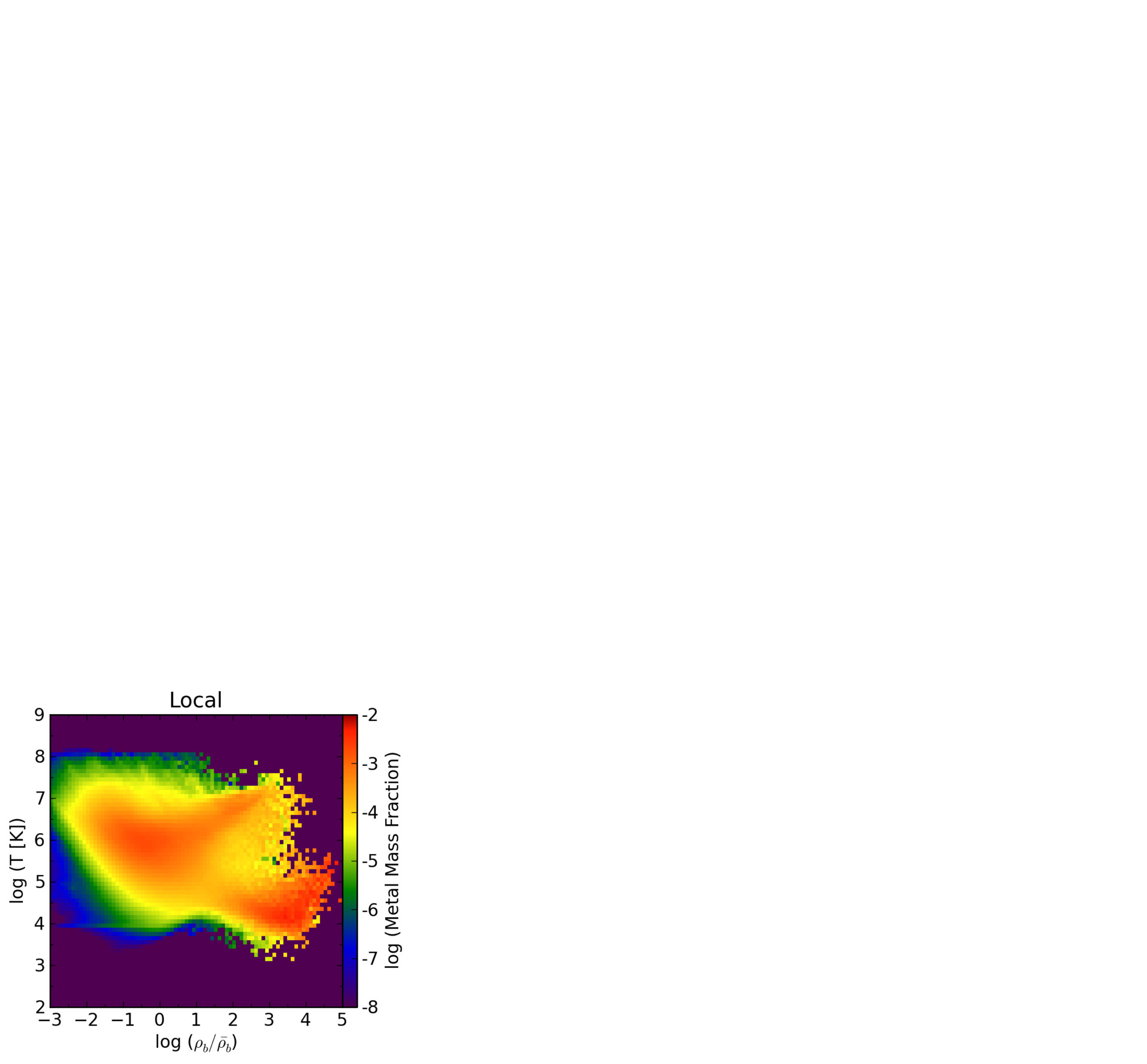}{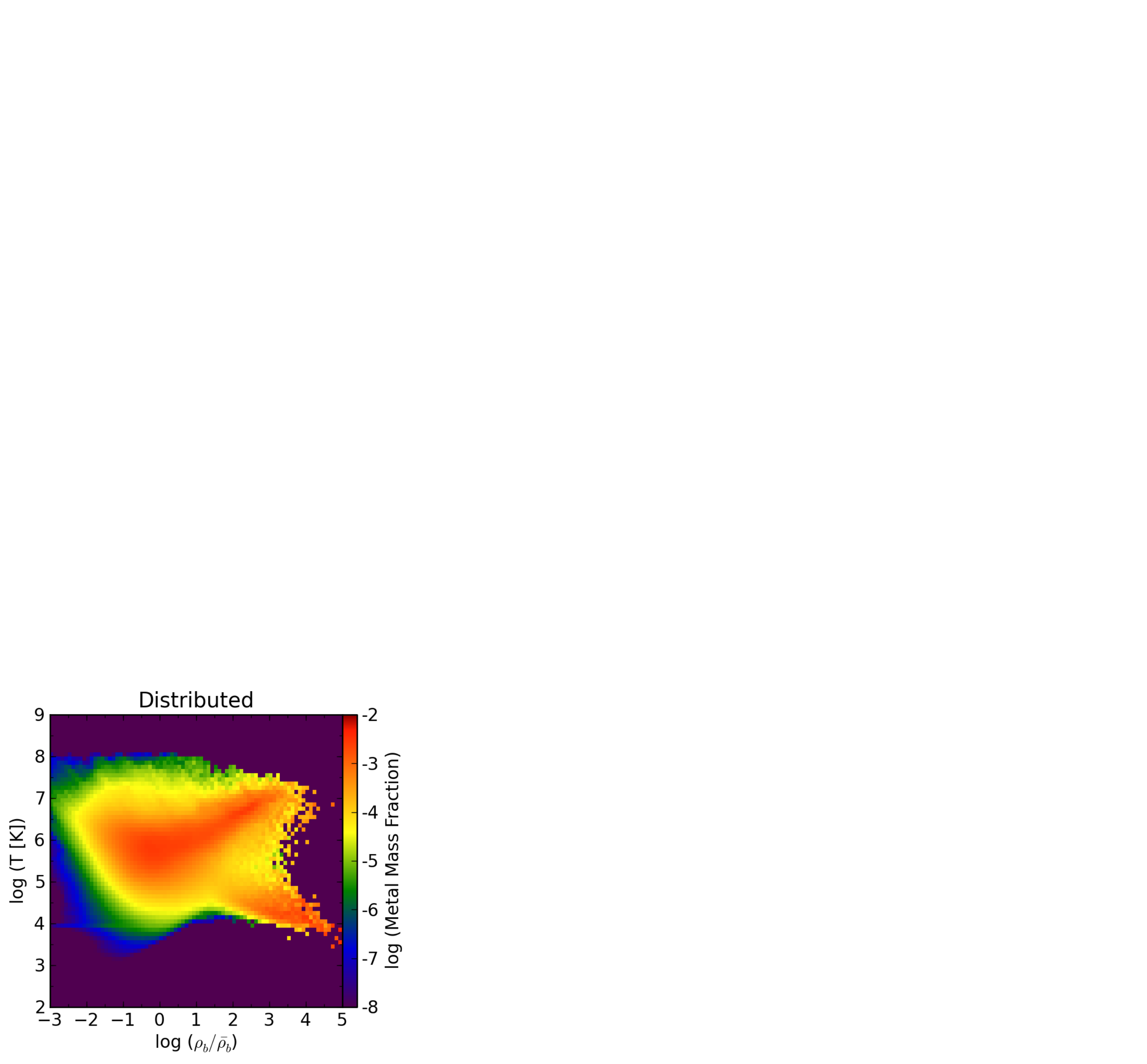}
  \caption{Fraction of total gas-phase metal mass in two-dimensional
    bins of overdensity ($\rho_{b}/\bar{\rho_{b}}$) and temperature
    for runs 25\_768\_1 (left) and 25\_768\_2 (right) at $z = 0$.  The
    primary regimes of metal population are the galaxies and the
    WHIM.  Although the total amount of gas-phase metals produced is
    roughly equivalent between the two runs, the amount of metals in
    stars is much higher for run 25\_768\_1.  The fraction of the
    total metal mass in gas (not stars) is 25\% for run 25\_768\_1 and
    45\% for run 25\_768\_2.}
 \label{fig:T_OD_Metals_25}
\end{figure*}

For the highest resolution 25 Mpc/$h$ simulations with normal metal
yield, the run with local feedback forms approximately
66\% more stars, and has essentially the same proportion of metals, 
compared to the run with distributed feedback, yet it produces 38\% 
fewer \OVI\ absorbers with \NOVI\ $ \ge 10^{12}\ \mathrm{cm}^{-2}$.  
How is this possible?  To answer this question, we examine the distribution
of gas-phase baryon (Figure \ref{fig:T_OD_Baryons_25}) and metal
mass (Figure \ref{fig:T_OD_Metals_25}) in bins of baryon overdensity
and temperature for runs 25\_768\_1 and 25\_768\_2.  In runs
25\_768\_1 and 25\_768\_2, $\sim75$\% and 85\% of the total
baryon mass is in the gas-phase, as opposed to in stars.  The distribution
of baryons in the two simulations is qualitatively similar, but there
are subtle differences.  The simulation with local
feedback has a significant population of warm, high density baryons ($T 
\lesssim 10^{5}$ K, $\rho_{b}/\bar{\rho_{b}} \gtrsim 10^{4}$), 
not present in the simulation with distributed feedback.  Instead,
the distributed-feedback simulation shows an increased amount of
hotter, slightly less dense baryons ($T\sim 10^{7}$ K,
$\rho_{b}/\bar{\rho_{b}} \sim 10^{2-3}$), and more
baryons in lower density WHIM as well.  Recall from Figure
\ref{fig:mass_fraction_25_768} that the local feedback run
has a WHIM mass fraction of 34\% at $z = 0$, compared to 44\% in the
distributed-feedback run.  

When considering the distribution of metals, the two simulations appear 
even more distinct.  A significant portion of the gas-phase metals in the local 
feedback run exist in the warm, high-density gas that is not present in 
the distributed-feedback run.  These metals are newly created by stars, yet 
unable to escape their points of origin due to over-cooling.  
While the two runs produce roughly the same number of gas-phase
metals, the run with local feedback produces significantly more metals
in stars.  Defining $\Omega_{metals, *}$ and $\Omega_{metals, gas}$ to
be the total amount of metals in stars and gas with respect to closure
density of the universe, we find $\Omega_{metals, *} =
1.17\times10^{-3}$ and $\Omega_{metals, gas} = 3.93\times10^{-4}$ for
the run with local feedback and $\Omega_{metals, *} =
4.81\times10^{-4}$ and $\Omega_{metals, gas} = 3.89\times10^{-4}$ for
the run with distributed feedback.  In contrast, the
distributed-feedback simulation shows far more metals in the WHIM 
phase.  To be exact, of the metals not in
stars, 44\% by mass are in the WHIM phase in the local
simulation, whereas 73\% are located in the WHIM in the distributed
simulation.  Differencing the two panels of Figure
\ref{fig:T_OD_Metals_25} reveals that the greatest increase in metal mass,
from the local to the distributed simulations, lies on an
adiabat extending from $T\sim 10^{7}$ K and $\rho_{b}/\bar{\rho_{b}}
\sim 10^{3}$ directly into the heart of the WHIM.  These are metals
outflowing from galaxies into the IGM, cooling adiabatically as they expand into
underdense gas.  In summary, even though the local feedback
simulation produces more metals, most of them become locked up in
their host galaxies, unable to escape into the IGM.  In contrast, the
distributed-feedback simulation creates overall fewer metals, but it
transports them into the WHIM far more efficiently.

\subsubsection{Does \ion{O}{6} Trace the WHIM?} \label{sec:trace_whim}

\begin{figure*}
  \plottwo{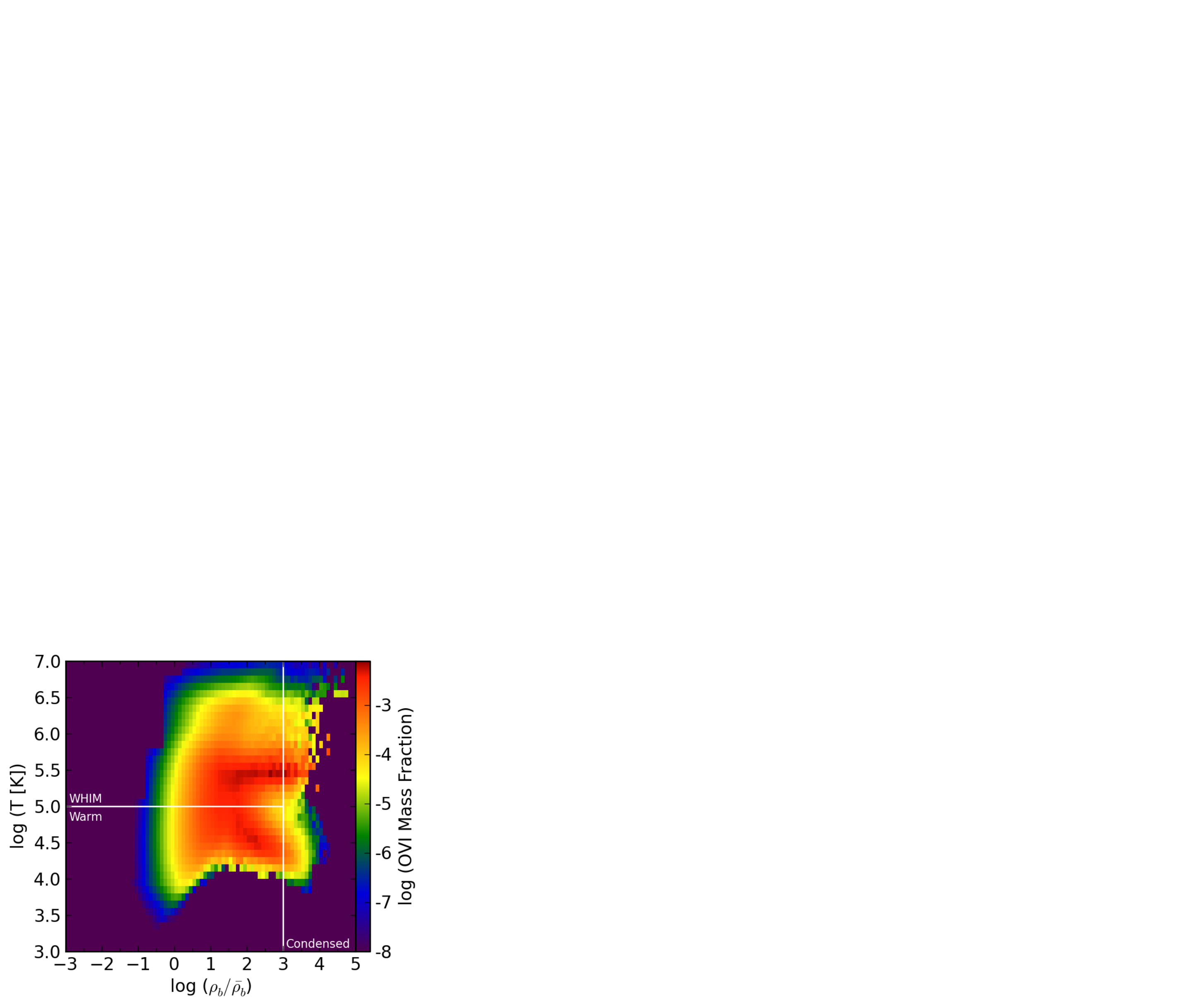}{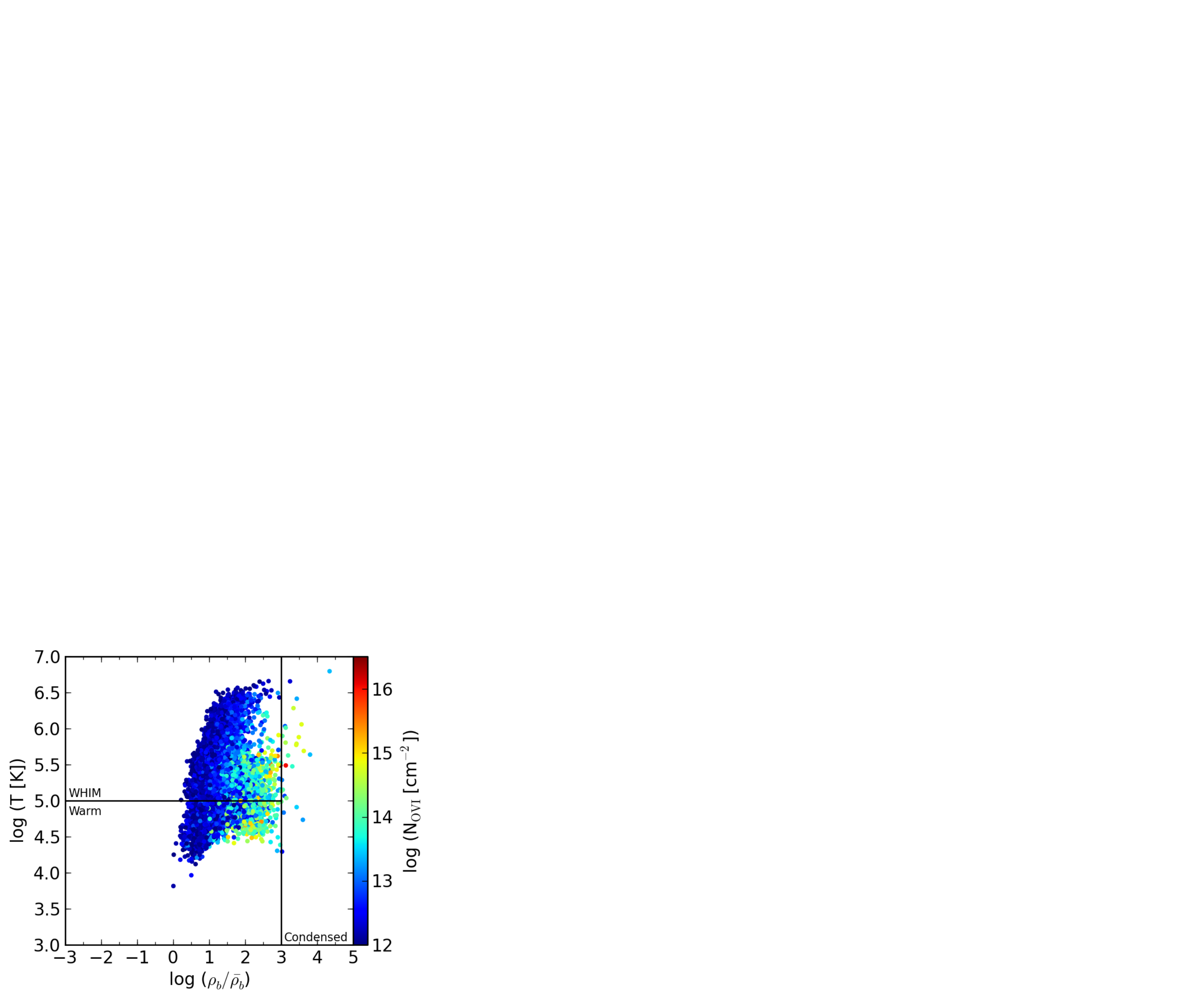}
  \caption{Left: fraction of total \ion{O}{6} mass in two-dimensional
    bins of baryon overdensity ($\rho_{b}/\bar{\rho_{b}}$) and temperature for run
    50\_1024\_2 at $z = 0$.  Right:  baryon overdensity and
    temperature associated with each synthetic \OVI\ absorber in run
    50\_1024\_2 (right panel of Figure \ref{fig:dndz_OVI_50}).  
    Colors correspond to the column density of the absorber.  In each
    panel, solid lines delineate divisions between the warm,
    WHIM, and condensed phases, as described in \S\ref{sec:baryon_phase}.}
 \label{fig:T_OD_OVI_Cloudy_50_2}
\end{figure*}

In order to answer this question, we plot the distribution of \OVI\ 
mass from the C+P model in bins of temperature and baryon overdensity
for run 50\_1024\_2 at $z = 0$ in the left panel 
of Figure \ref{fig:T_OD_OVI_Cloudy_50_2}.  From this, we can see
that the impression given by Figure \ref{fig:absorber_stats_25} that the
mean temperature of \OVI\ absorbers is $\sim 10^{5}$~K is 
misleading.  Instead, it appears that \OVI\ has a bimodal
distribution in temperature, with populations existing at
$\sim10^{4.5}$~K (mostly photoionized) and at $\sim10^{5.5}$~K (primarily
collisionally ionized).   We do not see any significant fraction of \OVI\ at 
$T \sim 15,000$ K, as claimed by
\citet{2009MNRAS.395.1875O}.  For the simulation shown, the fractions
of total \OVI\ mass are: 55\% in the WHIM, 37\% in the
warm phase, and 8\% in the condensed phase.  The fraction of \OVI\ in
the hot phase ($T > 10^{7}$~K) is negligible.  

Given these statistics, the next question is:
\textit{Are these phases proportionally represented by the \ion{O}{6} absorbers?}  
In the right panel of Figure
\ref{fig:T_OD_OVI_Cloudy_50_2}, we plot each one of the synthetic
\ion{O}{6} absorbers created for run 50\_1024\_2, using the C+P model, on
the plane of baryon overdensity and temperature with which they are
associated, coloring each absorber according to column density.  With
respect to the total column density, 59\% of the synthetically observed
\ion{O}{6} is in the WHIM, with 36\% in the warm phase and 5\% in the
condensed phase.  Given the somewhat arbitrary definition of the
condensed phase, the calculated fraction of \OVI\ in this
phase is not likely to be a very robust quantity.  If we consider only
\OVI\ absorbers with column 
densities greater than 10$^{13}\ \mathrm{cm}^{-2}$, closer to the observable 
range of \OVI\ with HST and FUSE, the fraction in the WHIM decreases to 57\%, 
the warm fraction increases to 37\%, and the condensed fraction increases
to 6\%.  This indicates that there is little biasing of the
observed \OVI\ away from the true \OVI\ distribution.  In contrast,
the simulations of \citet{2010arXiv1007.2840T} show a similar 
distribution of \OVI\ mass (bottom panel of Figure 3 in that work),
yet do not produce a significant number of 
\OVI\ absorbers in the warm phase.

\section{Discussion and Summary} \label{sec:summary}

We have performed a series of cosmological simulations designed to study
the evolution of the WHIM and its observability in low-redshift \OVI\ 
absorption lines.  The primary goals of this work are to find an
optimal parameterization of the processes of star formation and
feedback and to understand the extent to which the results of these
simulations are converged and can be trusted.  We investigated two
methods of depositing stellar feedback onto the simulation grid: 
(1) injecting all of the gas, metal, and thermal energy created by
a star particle into the single cell in which it is located
(local feedback); (2) distributing this feedback evenly over
the central cell and its 26 nearest neighbors (distributed feedback).
At low redshift ($z \lesssim 1$), the local-feedback
simulations show star formation rates (SFRs) much higher than observed.  
This appears to be the result of over-cooling that occurs when the
injection of thermal energy into a single grid cell results in
an unphysically high cooling rate.  We find that using the
distributed-feedback method significantly lowers the low-redshift SFR
and provides a much better fit to observations.  For both
feedback models, the SFR appears converged with respect to spatial and
mass resolution at low redshift.  At high redshift, the simulations are far from
converged, even though the SFRs of the highest-resolution runs are 
not in significant disagreement with observations, however unreliable
these observations may currently be.  Nevertheless, the 
final mass in stars is reasonably well converged, as the bulk of cosmic 
time occurs at low redshift.

The choice of feedback method has a significant impact on the fraction of
baryons in the WHIM.  For our highest-resolution simulations, there is
a 10\% difference in WHIM fraction at $z = 0$, with $\sim 35\%$ of
baryons in the WHIM in the local-feedback simulation and
$\sim 45\%$ in the distributed-feedback simulation.  However, these
results are not totally converged.  There is a significant lack of
convergence in the WHIM fraction for the 25 Mpc/$h$ box simulations with
local feedback.  The distributed-feedback simulations 
appear much closer to convergence on the WHIM fraction.  However,
those runs still show a slight increase in condensed gas and a decrease
in distributed gas with increasing resolution.  It is encouraging
that the baryon fractions for the distributed-feedback
simulations of both box sizes are in good agreement with each
other.  In slight contrast to other work on this subject, e.g.,
\citet{2006ApJ...650..560C, 2001ApJ...552..473D}, we do not see the
fraction of WHIM gas continuing to rise at $z = 0$.  Instead, all our
simulations show the WHIM fraction to peak at $z \sim 0.5$, then
decrease by a few percent by the present day.  As noted above in
\S\ref{sec:baryon_phase_feedback}, a small number of simulations in
other works have shown a decline in WHIM fraction at late time.
However, none of those studies see this phenomenon to be as ubiquitous
as we do.  We find that the decline in WHIM is
offset by a leveling off of the warm gas fraction that is decreasing
until $z \sim 0.5$ and a continual increase in the condensed
fraction.  By measuring the flux of baryons from one phase
to another, we see that the universe undergoes an epoch of heating
fueled by the formation of structure, followed by an epoch of adiabatic
cooling when the Hubble flow begins to accelerate due to the influence
of $\Omega_{\Lambda}$.  Encouragingly, the redshift at which
$\Omega_{\Lambda}$ becomes dominant ($z \sim 0.4$, with the
cosmological parameters used here) is consistent with when we observe
the WHIM fraction starting to decrease.  Radiative cooling has two
effects.  The total amount of condensed gas is increased by the
addition of radiative cooling, coming at the expense of WHIM gas.
However, radiative cooling, primarily from metals, also enables dense gas to
recool after being heated by stellar feedback, allowing it to continue
to form stars and inject heat into the IGM.

We used our highest-resolution simulations to create populations
of synthetic \OVI\ absorbers.  We compared these with the observed
number density per unit redshift ($d{\cal N}/dz$) of \OVI\ absorbers from 
$z = 0$ to $z = 0.4$ as measured by \citet{2008ApJ...679..194D}.  We 
experimented with two methods for calculating the fraction of oxygen in \OVI:  
the CIE model, which assumes collisional ionization equilibrium, and the C+P
model, which adds photoionization from the UV background of
\citet{2001cghr.confE..64H}.  Overall, the distributed-feedback
simulations, in conjunction with the C+P model for computing
$f_{\rm OVI}$, provide the best match to the observations.  The C+P model
provides an exceptional fit for low column density absorbers with 
\NOVI\ $ \lesssim 10^{14}\ \mathrm{cm}^{-2}$, but it slightly over-predicts the
number of higher column density absorbers, although still within the
error for the distributed-feedback simulations.  The CIE model fits
the $d{\cal N}/dz$ of the higher column density (\NOVI\ $ \gtrsim 10^{14}$
$\mathrm{cm}^{-2}$) absorbers better than the C+P model.  Previous studies have
also noted that low column density absorbers are better matched with
photoionization models, while high column density absorbers are better
matched with collisional ionization models \citep{2001ApJ...559L...5C,
  2001ApJ...561L..31F, 2003ApJ...594...42C}.
\citet{2005MNRAS.359..295F} used relatively simple arguments to
predict the maximum column density of characteristic \OVI\ absorbers
to be slightly over $10^{14}$ cm$^{-2}$.  In nearly all of our models,
we observe the $d{\cal N}(> $\NOVI$)/dz$ of \OVI\ to drop below unity
at roughly similar column densities.  
\citet{2006ApJ...650..573C} noted, as do we, that there is some
dependence of the number of low column density absorbers on feedback
method, with distributed-feedback methods producing more than 
local-feedback methods.  This difference is much more significant when using
the CIE model to calculate $f_{\rm OVI}$.  As is shown in Figure
\ref{fig:projections_50_2}, the CIE model produces large amounts of
\OVI\ at great distances from collapsed structures, precisely where
distributed-feedback methods excel over local feedback.  On the other
hand, the C+P model limits \OVI\ to regions much closer to galaxies
where the choice of feedback model is less likely to make such a
difference.  Given that the C+P model appears to be a much better fit
to the observed number density of low column density absorbers, the 
prospects for constraining the properties of galactic outflows with
\OVI\ may be limited.  In ionization equilibrium with the C+P model,
collisional ionization is dominant for 
$n \gtrsim 10^{-2}\ \mathrm{cm}^{-3}$ (Figure \ref{fig:ion_balance}).
However, the mean baryon density associated with \OVI\ absorbers with
the high column density \OVI\ absorbers is between
$\sim4\times10^{-6}$ cm$^{-3}$ and $\sim4\times10^{-4}\
\mathrm{cm}^{-3}$, or overdensities ranging from $\delta_H \approx$
23--2,300 at $z = 0$.   Future simulations that incorporate radiation
transport will likely be required to fully understand why the
influence of photoionization appears to be reduced in these regions.

The average physical conditions sampled by the two \OVI\ ionization models
are quite different, especially for low column densities.  The highest
column density absorbers, with \NOVI\ $ \sim 10^{15}\ \mathrm{cm}^{-2}$, 
have similar properties in both models and typically result from the rare combination 
of high physical density, high metallicity, and high $f_{\rm OVI}$.  In general, 
\OVI\ absorbers from the CIE model have higher associated baryon densities, 
lower metallicities, higher temperatures, and higher $f_{\rm OVI}$
than those in the C+P model.  In a later
paper, we will investigate the implications of the covariance in these parameters
for the (\OVI-based) IGM baryon census.  While \OVI\ absorbers from
the CIE model probe temperatures in a narrow range around the peak in $f_{\rm OVI}$
($\sim 300,000$ K), the average \OVI\ absorber temperature in the
C+P model is much closer to $100,000$ K.  This is noteworthy, as it
differs significantly from the recent simulations of
\citet{2009MNRAS.395.1875O} who find \ion{O}{6} absorbers to have average
temperatures of $\sim 15,000$ K.   Although the mean absorber
temperature is $\sim 100,000$ K in the C+P model, a closer inspection
reveals two distinct populations of \OVI\ at $\sim 30, 000$ K and
$\sim 300,000$ K.  We find that 55\% of all \OVI\  is in the WHIM,
37\% is in warm gas, and 8\% is in condensed gas.  These
proportions, which come from summing the total \OVI\ mass in the
simulation box at  $z = 0$, are reasonably well represented by the sample
of synthetic \OVI\  absorbers, with only a few percent bias toward the WHIM.

We acknowledge a number of caveats to our current work.  Radiation 
transport and ionizing sources are not explicitly present in our simulations, nor do we
include any ionizing sources or spatially varying radiation background.  The radiation
background is treated by altering the H/He photoionization rates in a
time varying, yet spatially uniform manner.  The addition of
radiation transport and ionizing sources may change the results
significantly, given that the \OVI\ absorbers are 
observed to be clustered around filaments \citep{2006ApJ...641..217S, 
2009ApJS..182..378W} and thus statistically nearer to 
sources of ionizing radiation.  We also do not track the non-equilibrium evolution 
of $f_{\rm OVI}$, as has been done by \citet{2006ApJ...650..573C} and
\citet{2006PASJ...58..641Y}.  While \citet{2006ApJ...650..573C} showed that
non-equilibrium calculations of $f_{\rm OVI}$ can alter the $d{\cal N}/dz$ 
distribution in \OVI\ column density, we were still able to fit the observations 
remarkably well with the C+P model.  In future work, we will include a
non-equilibrium calculation of the evolution of ionization fractions
for oxygen and other heavy elements.

We have shown that the number of \OVI\ absorbers depends sensitively on the 
metal yield from stars, but we have not included feedback processes that release 
metals in varying abundance patterns.  For instance, 
\citet{2008MNRAS.387.1431D} consider the feedback from Type Ia 
and Type II supernovae separately.  This also changes the rate of
injection of thermal energy.  The parameters we have chosen
for our star formation and feedback methods reflect our assumption that
stars form with a Salpeter IMF at all times, which is not necessarily
true.  \citet{2008MNRAS.385..687W} have pointed out 
inconsistencies between the observed star formation rate and the
observed stellar density; the observed SFR
implies a much higher stellar density than observed.  Since we
are able to match the observed evolution of SFR, our simulations
are also producing  too many stars.
\citet{2008MNRAS.385..687W} claim that this problem could
be solved if the stellar IMF was more top-heavy at high
redshift ($z \gtrsim 3$), an idea also supported by
\citet{2007ApJ...664L..63T}.  Other recent works have cited 
observational discrepancies that may be potentially solved with an
evolving IMF with characteristic mass increasing with redshift, e.g.,
\citet{2008MNRAS.385..147D, 2008ApJ...674...29V}.  
\citet{2009ApJ...691..441S} have shown that
the high-redshift CMB can suppress fragmentation in star forming
clouds, leading to an IMF that evolves with cosmic time and creates
additional massive stars in the past.  A varying IMF would have a
significant effect on the creation and dispersal of metals in the
early universe, and may be sufficiently important to consider in studies of
the enrichment of the IGM.

In subsequent papers, we will address the observability of the WHIM in
\OVII\ and \OVIII\, and other important elements, such as C, N, and
Ne.  We will also investigate other radiation backgrounds and their
influence on photoionized absorbers.  The slope of \dndz\ at low
column density is sensitive to the shape and intensity of the
radiation background, offering a means of constraining this background
with simulations.  In the very near future, it will be feasible to
include radiation transport in these simulations.  We will 
combine radiation transport hydrodynamics simulations with non-equilibrium
treatment of ionization balance in the metal species to study the
effect of local, non-uniform radiation on WHIM observables.

\acknowledgments

We are grateful to the anonymous referee for comments that 
considerably strengthened the paper.  
We thank Charles Danforth, Brian Keeney, Devin Silvia,
Sam Skillman, and Matthew Turk for helpful discussions, Andrew
Hopkins for kindly providing the data from
\citet{2006ApJ...651..142H}, and Stephen Wilkins for assisting us with
data from \citet{2008MNRAS.385..687W}.  This work was supported by the
Astrophysics Theory Program at the University of Colorado (NASA grant
NNZ07-AG77G and NSF grant AST-0707474).  EJH was
supported by NSF AAPF grant AST-0702923 and HST grant AR-12131.  BWO
was supported in part by NASA ATFP grant NNX09-AD80G.  
Computations described in this work were performed using the \texttt{Enzo} code
developed by the Laboratory for Computational Astrophysics at the
University of California in San Diego (http://lca.ucsd.edu) and by a
community of independent developers from numerous other institutions.
The \texttt{YT} analysis toolkit was developed primarily by Matthew Turk with
contributions from many other developers, to whom we are very
grateful.  The simulations were performed on NICS Kraken and TACC
Ranger, with computing time from Teragrid allocations TG-AST090040 and 
TG-AST100004.

\end{document}